\documentclass[
 reprint,
 amsmath,
 amssymb,
 aip
]{revtex4-2}
\usepackage[ruled,lined]{algorithm2e}
\usepackage[capitalise]{cleveref}
\usepackage{mathdots}
\usepackage{tikz}
\usepackage{pgfplots}
\usepackage{graphicx}
\usepackage{dcolumn}
\usepackage{bm}
\usepackage{url}

\begin{document}


\title{A fast, dense Chebyshev solver for electronic structure on GPUs}
\author{Joshua Finkelstein}
\email{jdf@lanl.gov}
 \affiliation{Theoretical Division, Los Alamos National Laboratory}
\author{Christian F. A. Negre}
\email{cnegre@lanl.gov}
 \affiliation{Theoretical Division, Los Alamos National Laboratory}
\author{Jean-Luc Fattebert}
 \email{fattebertj@ornl.gov}
\affiliation{
Computational Sciences and Engineering Division,  Oak Ridge National Laboratory
}

\date{\today}

\begin{abstract}
Matrix diagonalization is almost always involved in computing the density matrix needed in quantum chemistry calculations. In the case of modest matrix sizes ($\lesssim$ 5000), performance of traditional dense diagonalization algorithms on modern GPUs is underwhelming compared to the peak performance of these devices. This motivates the exploration of alternative algorithms better suited to these types of architectures. We newly derive, and present in detail, an existing Chebyshev expansion algorithm [W. Liang et al, J. Chem. Phys. 2003] whose number of required matrix multiplications scales with the square root of the number of terms in the expansion. Focusing on dense matrices of modest size, our implementation on GPUs results in large speed ups when compared to diagonalization. Additionally, we improve upon this existing method by capitalizing on the inherent task parallelism and concurrency in the algorithm. This improvement is implemented on GPUs by using CUDA and HIP streams via the MAGMA library and leads to a significant speed up over the serial-only approach for smaller ($\lesssim$ 1000) matrix sizes. Lastly, we apply our technique to a model system with a high density of states around the Fermi level which typically presents significant challenges.
\end{abstract}

\maketitle

\section{Introduction}\label{sec:intro}
Material science relies on the accurate calculation of electronic structure for computing material properties. Density matrix-based quantum chemistry methods, such as Density Functional Theory (DFT), and tight-binding based DFT, are standard approaches whereby the electron density of a chemical system is calculated, and from which ground-state quantum observables can then be obtained. In choosing a basis set to represent the electronic structure, for example a local basis of atomic orbitals, the problem becomes finite dimensional and we become concerned with calculating the single-particle density matrix, which we simply refer to as the density matrix (DM). The full DM calculation requires the self-consistent solution of a quantum-mechanical eigenvalue problem which is often the major bottleneck for all DM-based quantum chemistry methods. 

Perhaps the most straightforward approach to calculating the DM is to simply diagonalize the Hamiltonian matrix, at each self-consistent iteration step, and construct the DM from the eigenstates and energies. However, various other algorithms exist which avoid diagonalization by calculating an approximation to the DM through an expansion of the Fermi operator \cite{ANiklasson02, ANiklasson03}. In this communication, our aim is to approximate the DM using a Chebyshev polynomial expansion. Chebyshev expansions of the Fermi operator have been proposed in the context of linear scaling methods \cite{GoedeckerTeter1995,Baer1997} which sought to utilize the inherent sparsity of the Hamiltonian and DMs. While primarily developed for localized basis sets, these expansion techniques have been shown to be useful for wavefunction based DFT solvers too \cite{AaronsSkylaris2018, MOHR201864}. Chebyshev polynomial expansions have also been proposed and used as filters to avoid computing explicit eigenvectors in wavefunction based DFT \cite{Zhou2006,ALEMANY2007}, as well as in the context of stochastic DFT, especially when matrices are too large to be represented explicitly \cite{AWhite20, RBaer2013, YCytter2018, VSharma23}. Chebyshev expansions of the DM are also used with insulators and are still being developed and improved for that purpose \cite{MNguyen2022}.

Our focus is on algorithms that are well-suited toward GPU implementations, which are typically those that involve dense matrix-matrix multiplications. For example, in electronic structure calculations where systems exhibit a large electronic energy gap around the Fermi level, expansion algorithms based on matrix-matrix multiplications have proven to be highly efficient and accurate. Specifically, a dense matrix implementation of the second-order spectral projection (SP2) algorithm \cite{ANiklasson02} was shown to be very competitive when compared to dense diagonalization on Nvidia V100 GPUs for matrices smaller than $4000 \times 4000$ \cite{CoPA2021}. For larger matrix sizes (e.g. $20000 \times 20000$), SP2 has been shown to handily beat dense diagonalization when using new AI-based hardware such as Tensor cores from Nvidia \cite{JFinkelstein21}. Other density matrix purification methods were also shown to be extremely performant in a distributed context using Tensor Processing Units \cite{Pederson2023}. 

While some electronic structure problems can be quite large, domain scientists are often limited to solving problems of more modest sizes for which a very fast time-to-solution --- a few seconds or less --- can be achieved. This is specifically the case of quantum molecular dynamics where an electronic structure problem needs to be solved at each timestep. This leads to calculations that tend to push towards the strong scaling limit and use as many compute resources (nodes, cores) as possible. In wavefunction based approaches, where electronic wavefunctions are represented on a real-space mesh or with a planewave basis set, distribution of these wavefunctions over many cores and nodes can be used to significantly reduce the time-to-solution. However, even in this case, and despite powerful GPUs, diagonalization is still usually the bottleneck for computing the electronic structure. In the aforementioned case the matrix to be diagonalized is the dense Hamiltonian matrix in the subspace spanned by the wavefunctions \cite{LUPOPASINI2020}. 

We are motivated by matrices of size $N \times N$ where $N$ varies in the range of 500 to 5000 and want to determine how we can more efficiently use available GPU resources. This is the range of problem sizes for which time-to-solution tends to be a limitation for molecular dynamics, and for which matrix operations cannot take advantage of sparsity. This is particularly true for metallic systems in which the DM is fully dense. Dense diagonalization implementations for GPUs, such as the one offered in Nvidia's cuSolver \cite{cusolver}, have already been shown to have limited performance for modestly sized matrices \cite{CoPA2021}. 

It was back in the 1970's that Patterson and Stockmeyer \cite{Patterson1973} first showed that any generic polynomial $P$ in powers of $x$ with $L$ terms could be rewritten in such a way that only $\sim 2\sqrt{L}$ multiplications in $x$ are needed to evaluate $P(x)$. About 30 years later, in the early 2000's, Liang et al. \cite{Liang2003,Liang2004} showed that this same idea could be applied directly to Chebyshev polynomial expansions of the Fermi operator and not just to polynomials expressed in the standard monomial basis. Here, we demonstrate the power of this approach on GPUs and present a parallelized version targeting small or moderately sized dense matrices.  This parallelized implementation utilizes modern GPU concurrency, in particular CUDA and HIP streams, making it a useful technique for the new class of heterogeneous architectures now available on today's peta and exascale machines. 

In Section \ref{sec:background} we briefly discuss some background on the density matrix and diagonalizations compared to matrix multiplications on GPUs. We also discuss the Chebyshev polynomial expansion used in place of diagonalization and elaborate the $\mathcal{O}(\sqrt{L})$ technique used to reduce the number of matrix multiplications needed. Section \ref{sec:implementation} will explain our implementations using MAGMA \cite{magma}, as well as discuss our technique to enable task parallelism and GPU concurrency. Lastly, in Section \ref{sec:results}, we examine the application of this technique to a model Hamiltonian and discuss the resulting speed ups and time-to-solution observed with matrix sizes considered to be small for today's modern GPU devices.

\section{Algorithm} \label{sec:background}

\subsection{Density Matrix}

The most direct way to compute the density matrix, $D$, is by computing the eigenvectors $\{\mathbf v_i\}$ and eigenvalues $\{\varepsilon_i\}$ of the systems' $N \times N$ Hamiltonian matrix, $H$. If $E$ is the diagonal matrix with diagonal elements $E_{ii}=\varepsilon_i$, and $V$ is the  matrix made of columns $\mathbf{v_i}$ ordered so that $\varepsilon_i$ and $\mathbf{v_i}$ are eigenpairs of $H$, then $D$ is given by
\begin{equation} \label{eq:D from diag}
    D=Vf(E)V^T \;,
\end{equation}
where
\begin{equation}
    f(\varepsilon) = \frac{1}{1 + \exp(\beta( \varepsilon-\mu))} \;,
\label{eq:fermifunc}
\end{equation}
is the Fermi-Dirac distribution function, $\beta$ is the inverse electronic temperature and $\mu$ is the chemical potential, or Fermi level. 

\begin{figure}[ht]
\centering
    \begin{tikzpicture}
        \begin{loglogaxis}[width=.485\textwidth,height=2.5in,
                    xmin=400, xmax=20000,
                    ymin=1, ymax=50000,
                    legend pos=north east,
                    legend columns=2,
                    ylabel={Relative time-to-solution},
                    xlabel={Matrix size ($N$)},
                    scaled x ticks = false,
                    ymajorgrids=true,
                    grid style=dashed,
                    domain=100:1000,
                    legend entries={V100, EPYC 1 thread, MI250X, EPYC 4 threads}
                    ]
        \addplot[blue,mark=square*,line width=0.5] table {data/v100_diag_vs_mm.dat};
        \addplot[brown,mark=x, mark size = 3,line width=0.75] table 
        {data/epyc1_diag_vs_mm.dat};
        \addplot[red,mark=triangle*, line width=0.75] table {data/mi250x_diag_vs_mm.dat};
        \addplot[orange,mark=*, line width=0.75] table 
        {data/epyc4_diag_vs_mm.dat};
        \end{loglogaxis}
    \end{tikzpicture}
\caption{Relative time-to-solution for a dense diagonalization compared to a dense matrix-matrix multiplication for different GPUs, CPUs and matrix sizes. Random square symmetric matrices are used.}
\label{fig:dsyevd}
\end{figure}
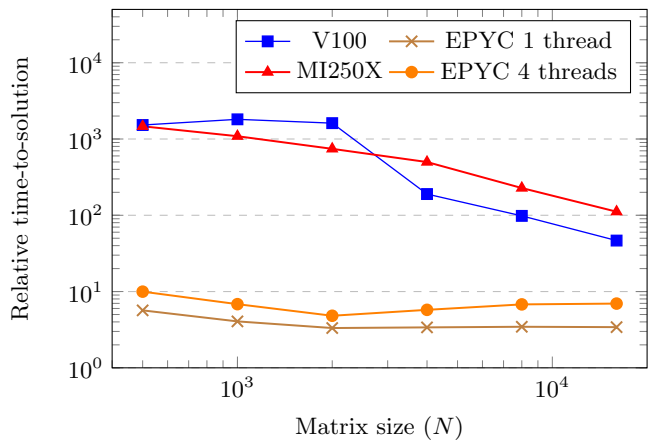

Implementing an efficient dense diagonalization algorithm to compute the eigenpairs $\{\varepsilon_i,\bf v_i\}$ is a difficult task compared to matrix-matrix multiplication, and, it is even more difficult to develop diagonalization algorithms that run efficiently on GPUs. While a dense diagonalization requires about the same number of operations as a dense matrix-matrix multiplication \cite{demmel97}, time-to-solution differs substantially.
Fig.~\ref{fig:dsyevd} shows the relative time-to-solution for dense diagonalization of a symmetric random matrix (including the eigenvectors computation) compared to a dense matrix-matrix multiplication on several computer architectures. The timings were obtained using the OpenBLAS library \cite{openblas} on CPUs and the MAGMA library \cite{magma,magma2014} on GPUs. The DSYEVD function, that is the \textit{divide and conquer} version of diagonalization, was used for all performance measurements. The timings for the aforementioned dense diagonalization were divided by the timings of a single dense matrix-matrix multiplication, using the DGEMM function to obtain the relative times \footnote{It is worth noting that ${\tt magma\_dgemm}$ and ${\tt magmablas\_dgemm}$ are different implementations within MAGMA, with the former utilizing vendor libraries and latter the intrinsic MAGMA version. In our implementations, we use ${\tt magma\_dgemm}$}. While that ratio remains below ten on a CPU, it goes up substantially on GPUs, in particular for matrices of moderate sizes ($N <$ 8000).

\subsection{Chebyshev expansions with $\mathcal{O}(\sqrt{L})
$ number of matrix multiplications}\label{sec:method+parallelization}
As mentioned in the introduction, instead of trying to compute $D$ directly, other algorithms have been proposed that calculate an approximation to $D$. Rather than diagonalizing $H$ as in \cref{eq:D from diag}, the construction of $D$ can equivalently be formulated as a direct application of the Fermi-Dirac function $f$ to the Hamiltonian matrix $H$,
\begin{align}
    f(H) = [I + \exp(\beta(H -\mu I))]^{-1} \label{eq:fermi} = D \;.
\end{align}
In this context, $f$ is often called the Fermi operator. We then seek to approximate $f$ through a Chebyshev expansion. It was shown previously\cite{Liang2003} that a Chebyshev expansion of length $L$, given by $\sum_{n=0}^L c_n T_n$, could be expanded using the ansatz
\begin{align}
\begin{split}\label{eq:ansatz}
    \sum_{n=0}^L c_n T_n  = \sum_{i=0}^{k-1} & d_{i,0} T_i + T_k 
                \bigg(\sum_{i=0}^{k-1} d_{i,1} T_i + T_k \bigg( \sum_{i=0}^{k-1} d_{i,2} T_i \\
                & + \cdots + T_k \bigg( \sum_{i=0}^{k-1} d_{i,{m-1}} T_i \bigg) \cdots \bigg)\bigg)\;, 
\end{split}
\end{align}
where it is assumed that $L = km-1$, with $k, m$ positive integers and $T_n$ are the $n$-th Chebyshev polynomials with $T_0=(H-\varepsilon_a)/(\varepsilon_b-\varepsilon_a)$ for $\varepsilon_a$ and $\varepsilon_b$ the lowest and highest eigenvalues of $H$, respectively. The coefficients $d_{i,j}$ can then be systematically determined by matching the terms on both sides of \cref{eq:ansatz}. The direct benefit of this form is that we only need to compute Chebyshev polynomials of order up to $k$, which requires just $k-1$ matrix multiplications. Once $T_0,\dots,T_k$ are known, an additional $m-1$ matrix multiplications are then needed to multiply $T_k$ with the Chebyshev polynomials inside each parenthesis to the right of each $T_k$ in \cref{eq:ansatz}. Note that the number of matrix multiplications is minimized when $k=m=\sqrt{L+1}$, the situation we consider in our numerical tests. The format in \cref{eq:ansatz} is also slightly different than what is described in Ref.~\onlinecite{Liang2003}. There, the summations can be of different lengths, whereas we assume them to all be the same. This is both a natural choice for our parallelized approach and greatly simplifies presentation and implementation. 

The right-hand side of \cref{eq:ansatz} can be expressed as 
\begin{align}\label{eq:ansatz2}
\begin{split} 
\sum_{i=0}^{k-1} d_{i,0}T_i + T_k \bigg( \sum_{i=0}^{k-1} &d_{i,1} T_i \bigg) + T_k^2 \bigg( \sum_{i=0}^{k-1} d_{i,2} T_i\bigg) + \\
 & \cdots  + T_k^{m-1}  \bigg( \sum_{i=0}^{k-1} d_{i,{m-1}} T_i \bigg)  \\
\end{split}
\end{align}
by multiplying through with $T_k$, so that \cref{eq:ansatz} can then be rewritten as 
\begin{align}\label{eq:ansatz3}
\begin{split} 
\sum_{n=0}^L c_n T_n =\sum_{j=0}^{m-1} \bigg( \sum_{i=0}^{k-1} d_{i,j}T_k^jT_i \bigg) \;.
\end{split}
\end{align}
To determine the expansion coefficients $d_{i,j}$ on the right hand side of \cref{eq:ansatz3}, we make use of the basic multiplicative identity for Chebyshev polynomials of the first kind 
\begin{align}
 2T_nT_m =T_{n+m} + T_{|n-m|} \label{eq:identity} \;,
\end{align}
for non-negative integers $n,m$. First we compute the expansions for the various powers of $T_k$. The first few expansions are
\begin{align}
\begin{split} \label{eq:example powers of Tk}
    T_k^2 &= \frac{1}{2}(T_{0}+T_{2k})\\
    T_k^3 &= \frac{1}{4}(3T_{k}+T_{3k})\\
    T_k^4 &= \frac{1}{8}(3T_0 + 4T_{2k} + T_{4k})  \;.
\end{split}
\end{align}
In general, non-negative powers of $T_k$ can be written in a Chebyshev basis as
\begin{align}\label{eq:power_of_Tk}
T_k^j=\sum_{\ell=0}^{j}  a_{j,\ell} T_{\ell \times k}\;,
\end{align}
and it can then be deduced that the expansion coefficients $a_{j,\ell}$, for the power $j$ term, are given by a recurrence, from the coefficients $a_{j-1,\ell}$ of the expansion for $T_k^{j-1}$, as
\begin{align} \label{eq:coeff recursion}
a_{j,\ell} = 
\begin{cases}
\tfrac{1}{2}a_{j-1,1}  &\text{if } \ell=0\\
a_{j-1,0} + \tfrac{1}{2} a_{j-1,2} &\text{if } \ell=1\\
\tfrac{1}{2} (a_{j-1,\ell-1} + a_{j-1,\ell+1}) &\text{if } 1< \ell <j\\
\frac{1}{2}a_{j-1,\ell-1} &\text{if } \ell=j 
\end{cases}
\end{align}
for $1<j$ and $0\le \ell \le j$, with the first few terms of the recurrence being $a_{0,0} = 1$, $a_{1,0} =0$ and $a_{1,1} = 1$. This calculation is the origin of the recurrence formula presented in Refs.~\onlinecite{Liang2003, Liang2004}. 
The $a_{j,\ell}$ so given, allow us to define the $m\times m$ matrix ${\bf A}$ with elements
\begin{align}
    ({\bf A})_{j\ell} &\equiv \begin{cases}
    a_{j,\ell} & \ell \le j\\
    0 & {\ell >j} 
\end{cases}\;,
\end{align}
for $0 \le  j,\ell \le m-1$. By construction, ${\bf A}$ is a lower-triangular matrix and has repeating sub-diagonals of zeros, which is a consequence of $T_k^j$ being a sum of only Chebyshev polynomials with $\ell$ the same parity as $j$ in \cref{eq:power_of_Tk}. To illustrate this, the example calculations in \cref{eq:example powers of Tk} can be used with $L=24$, $k=m=5$, so that in this case the matrix ${\bf A}$ would be 
\begin{align}
\label{eq:example A}
    {\bf A} = 
    \begin{pmatrix}
        1 & 0 & 0 & 0 & 0\\
        0 & 1 & 0 & 0 & 0 \\
        \tfrac{1}{2} & 0 & \tfrac{1}{2} & 0 & 0 \\
        0 & \tfrac{3}{4} & 0 & \tfrac{1}{4} & 0 \\
        \tfrac{3}{8} & 0 & \tfrac{4}{8} & 0 & \tfrac{1}{8}\\
    \end{pmatrix}\;.
\end{align}
This matrix ${\bf A}$ can be used to directly solve for the unknown $d_{i,j}$ coefficients. The expression on the right-hand side of \cref{eq:ansatz3} can be rewritten as 
\begin{align*}
    \sum_{j=0}^{m-1} \sum_{i=0}^{k-1}  d_{i,j} T^j_k T_i 
    & = \sum_{i=0}^{k-1} \bigg( \sum_{j=0}^{m-1}  d_{i,j} T^j_k \bigg)T_i \nonumber \\
    & = \sum_{i=0}^{k-1} \bigg( \sum_{j=0}^{m-1}  \bigg[ \sum_{\ell=0}^{j} d_{i,j} a_{j,\ell} T_{\ell\times k} \bigg] \bigg)T_i \nonumber \\
    & = \sum_{i=0}^{k-1} \bigg( \sum_{\ell=0}^{m-1}  \bigg[ \sum_{j=0}^{m-1} d_{i,j} a_{j,\ell}  \bigg] T_{\ell\times k} \bigg)T_i \nonumber \\
    & = \sum_{i=0}^{k-1} \bigg( \sum_{\ell=0}^{m-1} ({\bf d}_i^T  {\bf a}_\ell)  T_{\ell\times k} \bigg)T_i \;,
\end{align*}
where ${\bf d}_i$ is a vector of size $m$ with components $d_{i,j}$ with $j=0,\dots,m-1$, and ${\bf a}_\ell$ is the $\ell$-th column of ${\bf A}$. Using again \cref{eq:identity}, we have, for $1 \le \ell \le m-1$  and $1\le i \le k-1$,
\begin{align}
    T_{\ell\times k}T_{i} = \frac{1}{2}  (T_{\ell\times k+i}+ T_{\ell\times k-i}).
\end{align}
Also, if we set $i'=k-i$, we can write
\begin{align}
({\bf d}_i^T {\bf a}_\ell) T_{\ell\times k-i} &= ({\bf d}_{i}^T {\bf a}_\ell) T_{({\ell-1})\times k+k-i} \nonumber \\
&= ({\bf d}_{k-i'}^T {\bf a}_\ell) T_{({\ell-1})\times k + i'} \;,
\end{align}
so that $1\le i\le k-1$ implies $1\le i'\le k-1$, and
\begin{align}
\sum_{i=1}^{k-1} ({\bf d}_{k-i}^T {\bf a}_\ell) T_{({\ell-1})\times k+i} = \sum_{i=1}^{k-1} ({\bf d}_i^T {\bf a}_\ell) T_{\ell\times k-i}\;.
\end{align}
Therefore,
\begin{align*}
    \sum_{j=0}^{m-1} \sum_{i=0}^{k-1}  d_{i,j} T^j_k T_i 
    & = \sum_{\ell=0}^{m-1}  ({\bf d}_0^T {\bf a}_\ell)T_{\ell\times k} \\
    & + \sum_{i=1}^{k-1} ({\bf d}_i^T  {\bf a}_0 + \tfrac{1}{2}{\bf d}_{k-i}^T {\bf a}_1)T_i \\ 
    & + \frac{1}{2} \sum_{\ell=1}^{m-2} \sum_{i=1}^{k-1}  ({\bf d}_i^T {\bf a}_\ell + {\bf d}_{k-i}^T {\bf a}_{\ell+1} )T_{\ell\times k+i} \\ 
    & + \frac{1}{2} \sum_{i=1}^{k-1} ({\bf d}_i^T {\bf a}_{m-1}) T_{(m-1)\times k+i}\;.
\end{align*}

With this calculation, we have now fully expressed the right-hand side of \cref{eq:ansatz3} as a Chebyshev polynomial expansion, and these expansion coefficients can be matched to the original Chebyshev expansion coefficients on the left-hand side of \cref{eq:ansatz3}. Thus, to satisfy \cref{eq:ansatz3}, for each $i = 0, 1,..., k-1$ and $\ell = 0, 1, ..., m-1$, we require that the vectors ${\bf d}_i$ solve the equations
\begin{align}
\label{eq:formula for d0}
    c_{\ell\times k} &= {\bf d}_0^T {\bf a}_\ell  \;, 
\end{align}
and
\begin{align}
\label{eq:formula for d}
\begin{split}
    c_{\ell\times k + i} &= \begin{cases}
        {\bf d}_i^T {\bf a}_{0}  + \tfrac{1}{2}{\bf d}_{k-i}^T {\bf a}_{1} & \;, \ell = 0\\
        \tfrac{1}{2} {\bf d}_i^T {\bf a}_\ell  + \tfrac{1}{2}{\bf d}_{k-i}^T {\bf a}_{\ell+1}  & \;, 1\le \ell \le m-2 \\
        \tfrac{1}{2} {\bf d}_i^T {\bf a}_{m-1}  & \;, \ell = m-1
    \end{cases} 
\end{split}
\end{align}

These equations shown in \cref{eq:formula for d,eq:formula for d0} can be combined into a single upper-triangular linear system that solves for all $\ell$ and $i$ at once. In the Appendix, we work through the algebra which derives, step-by-step, the form shown below. Once expressed this way, all of the $d_{i,j}$ coefficients are solved for using back substitution. Defining ${\bf J}$ to be the $(k-1) \times (k-1)$ column-reversed identity, i.e. 
\begin{align}
    {\bf J} \equiv
    \begin{pmatrix}
        &   &          &          &  &    & 1  \\
        &   &          &          &  &  1 &    \\
        &   &          &  \iddots &  &    &    \\ 
        & 1 &          &          &  &    &    \\ 
     1  &   &          &          &  &    &    \\ 
    \end{pmatrix} \;,
\end{align}
we are then able to write 
\begin{align}
    \begin{bmatrix}
        c_{0}   \\
        c_{1}   \\
        \vdots  \\
        c_{k-1} \\
        \vdots \\
        c_{(m-1) \times k}\\
        c_{(m-1) \times k + 1}\\
        \vdots\\
        c_{(m-1) \times k + (k-1)}
    \end{bmatrix}  = {\bf X}
    \begin{bmatrix}
        d_{0,0} \\
        d_{1,0} \\
        \vdots \\
        d_{k-1,0} \\
        \vdots \\
        d_{0,m-1} \\
        d_{1,m-1} \\
        \vdots \\
        d_{k-1,m-1} 
    \end{bmatrix} 
\end{align}
\begin{widetext}
where ${\bf X}$ is the is the $km\times km$ block matrix given by
\begin{align} \label{eq:X}
    {\scriptsize
    \begin{bmatrix}
     & \ddots & \vdots & & & &  \vdots & & & & \vdots \\
    & a_{m-5,m-5} & & 0 & &  a_{m-3,m-5} &   & 0 &  & a_{m-1,m-5}&  & & \\
    & & \tfrac{1}{2}a_{m-5,m-5}{\bf I} & & \tfrac{1}{2}a_{m-4,m-4}{\bf J}  &   &  \tfrac{1}{2}a_{m-3,m-5}{\bf I} & &   \tfrac{1}{2} a_{m-2,m-4} {\bf J}& & \tfrac{1}{2}a_{m-1,m-5}{\bf I} & \\
    & & & a_{m-4,m-4} & & 0 &  &  a_{m-2,m-4} & & 0 & \\
    & & & & \tfrac{1}{2}a_{m-4,m-4} {\bf I} & & \tfrac{1}{2}a_{m-3,m-3}{\bf J} & & \tfrac{1}{2}a_{m-2,m-4} {\bf I} &  & \tfrac{1}{2}a_{m-1,m-3} {\bf J} \\
    & & & & & a_{m-3,m-3} & & 0 & & a_{m-1,m-3} & &\\
    & & & & & & \tfrac{1}{2} a_{m-3,m-3} {\bf I} & & \tfrac{1}{2} a_{m-2,m-2} {\bf J}  & & \tfrac{1}{2} a_{m-1,m-3} {\bf I} \\
    & & & & & & &a_{m-2,m-2} & & 0 &  & & \\
    & & & & & & & & \tfrac{1}{2} a_{m-2,m-2} {\bf I} & & \tfrac{1}{2} a_{m-1,m-1} {\bf J} \\
    & & & & & & & & & a_{m-1,m-1}  & \\
    & & & & & & & & & & \tfrac{1}{2} a_{m-1,m-1} {\bf I} 
    \end{bmatrix} }
\end{align}
\end{widetext}
with ${\bf I}$ the $(k-1) \times (k-1)$ identity matrix. The columns of ${\bf X}$ containing the ${\bf I}$ and ${\bf J}$ matrices are columns of block matrices and the columns containing only zeros and entries of the ${\bf A}$ matrix, $a_{j,\ell}$, are columns of scalars. Zeros are only written in the matrix entries containing scalars. The pattern of each row and column shown above continues until the first row with the exception that in the first row, the ${\bf I}$ blocks do not include the factor of $\tfrac{1}{2}$ as in the top of \cref{eq:formula for d}.

\section{Implementation and Numerical results}
\label{sec:implementation}
Our implementation is based on the MAGMA library \cite{magma2014,magma2020,magma} which offers performance and portability for dense linear algebra operations. Among the different architecture types supported by MAGMA, we find both Nvidia and AMD GPUs, which are the most widely used types. All numerical tests were done on Summit and Crusher at the Oak Ridge Leadership Computing Facility (OLCF), using MAGMA v2.6.1 and the Basic Matrix Library (BML) \cite{Bock2018}. Each Summit compute node has two 44-core IBM Power9 CPUs and six Nvidia V100 GPUs with a peak performance of 7.8 TFLOP/s in double-precision. Like Frontier, the current number one system on the top500 list of supercomputers, Crusher compute nodes have one 64-core AMD EPYC CPU and four AMD MI250X, each composed of two GPUs with a peak performance of 26.5 TFLOPS in double-precision. These GPUs allow for concurrent streams. A stream is a sequence of operations that execute in issue-order on the GPU. Operations executed in different streams may run concurrently. 

\subsection{Speed up over diagonalization}
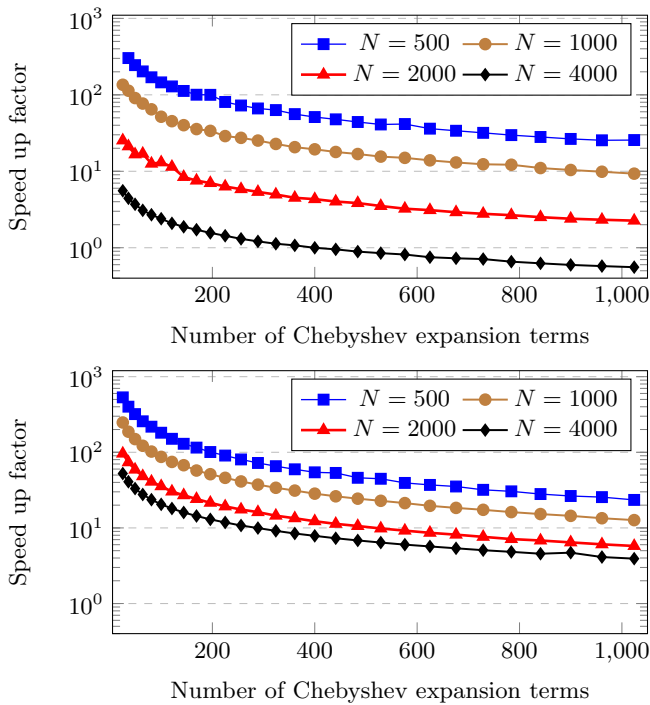
\begin{figure}
    \centering
        \begin{tikzpicture}
        \begin{semilogyaxis}[width=.485\textwidth,height=2.in,
                    xmin=5, xmax=1050,
                    ymin=0.4, ymax=1100,
                    legend pos=north east,
                    legend columns=2,
                    ylabel={Speed up factor},
                    xlabel={Number of Chebyshev expansion terms},
                    ymajorgrids=true,
                    grid style=dashed,
                    domain=100:1000,
                    legend entries={$N=500$, 
                    $N=1000$,
                    $N=2000$, 
                    $N=4000$}]
        \addplot[blue,mark=square*,line width=0.5] table {data/500x500_diag_vs_PS};
        \addplot[brown,mark=*, line width=0.75] table 
        {data/1000x1000_diag_vs_PS};  
        \addplot[red,mark=triangle*, line width=1.0] table {data/2000x2000_diag_vs_PS.dat};
        \addplot[black,mark=diamond*, line width=0.75] table {data/4000x4000_diag_vs_PS.dat};
        \end{semilogyaxis}
    \end{tikzpicture}
    \begin{tikzpicture}
        \begin{semilogyaxis}[width=.485\textwidth,height=2.in,
                    xmin=5, xmax=1050,
                    ymin=0.4, ymax=1200,
                    legend pos=north east,
                    legend columns=2,
                    ylabel={Speed up factor},
                    xlabel={Number of Chebyshev expansion terms},
                    ymajorgrids=true,
                    grid style=dashed,
                    domain=100:1000,
                    legend entries={$N=500$, 
                    $N=1000$,
                    $N=2000$, 
                    $N=4000$}]
        \addplot[blue,mark=square*,line width=0.5] table {data/crusher/500x500_diag_vs_PS-crusher.dat};
        \addplot[brown,mark=*, line width=0.75] table 
        {data/crusher/1000x1000_diag_vs_PS-crusher.dat};
        \addplot[red,mark=triangle*, line width=1.0] table {data/crusher/2000x2000_diag_vs_PS-crusher.dat};
        \addplot[black,mark=diamond*, line width=0.75] table {data/crusher/4000x4000_diag_vs_PS-crusher.dat};
        \end{semilogyaxis}
    \end{tikzpicture}
    \caption{(top) Speed up over DM construction with diagonalization on an Nvidia V100 GPU from using the $\mathcal{O}(\sqrt{L})$ Chebyshev method. The BML library is used for diagonalization (which calls cuSolver). Only a single GPU stream is used to compute the Chebyshev expansion. (bottom) Speed up over DM construction using diagonalization on an AMD MI250X GPU for a single stream. The MAGMA library is used for diagonalization.} \label{fig:speed up_diag}
\end{figure}

\Cref{fig:speed up_diag} shows the speed up obtained when using the $\mathcal{O}(\sqrt{L})$ Chebyshev polynomial expansion technique over diagonalization. The BML library is used for diagonalization on Nvidia GPUs, where BML's diagonalization function is essentially a wrapper around the cuSolver function ${\tt cusolverDnDsyevd}$. Only a single GPU stream is used here in order to emphasize that any performance increase is due exclusively to the more efficient $\mathcal{O}(\sqrt{L})$ Chebyshev methodology and not to concurrency. On the AMD GPU, the MAGMA diagonalization function ${\tt magma\_dsyevd}$ was used instead of the cuSolver function, and the speed up is shown in the bottom panel of \cref{fig:speed up_diag}. For the V100, the largest speed ups occur for smaller matrix sizes. When $N=4000$, we actually see a slow down relative to diagonalization for large enough expansions. At this point, diagonalization becomes more efficient. The bottom figure shows more modest speed up factors for the MI250X for smaller matrix sizes, but larger ones for for $N=2000$ and $N=4000$. We attribute these differences between the two GPUs to the more performant diagonalization available on the V100 with cuSolver. The speed ups on the MI250X are also less sensitive to matrix size. For a given size, the difference between a $500 \times 500$ and $4000 \times 4000$ matrix is only around an order of magnitude, whereas for the V100, it is around two orders of magnitude. Therefore on the MI250X, even for a more moderately sized $4000 \times 4000$ matrix, the Chebyshev expansion method still offers some advantage, even with many terms.

\subsection{Speed up with multiple GPU streams}
The expression on the right hand side of \cref{eq:ansatz} can be grouped using variables $S_j, j=0,\dots, m-1$ as 
\begin{align}
\begin{split}
    \underset{S_0}{\underbrace{\sum_{i=0}^{k-1} d_{i,0}T_i}} + &T_k 
                            \bigg(\underset{S_1}{\underbrace{\sum_{i=0}^{k-1}d_{i,1} T_i}} +  T_k \bigg( \underset{S_2}{\underbrace{\sum_{i=0}^{k-1} d_{i,2} T_i}} 
                            + \\
                            & \cdots + T_k \bigg( \underset{S_{m-1}}{\underbrace{\sum_{i=0}^{k-1} d_{i,{m-1}} T_i }}\bigg) \cdots \bigg)\bigg) \;.
\end{split}
\end{align}
The important observation to make is that each low-order Chebyshev expansion $S_j$ are independent from one another as they only depend upon the precomputed coefficients $d_{i,j}$ and the shared Chebyshev polynomials $T_i$. This introduces some opportunity for parallelism in the Chebyshev expansion calculation since all the $S_j$ terms can then be summed concurrently. 

Further parallelism can be obtained by recognizing that the Chebyshev polynomials $T_i$ themselves, $i=2,...,k-1$, need not be calculated in serial. Each $T_i$ can be calculated by making use of the Chebyshev polynomial multiplication identity in \cref{eq:identity}. This comes down to the fact that from $T_0$ and $T_1$ we obtain
\begin{align}\label{eq:cheby1}
    T_2 = 2T_1T_1 - T_0\;,
\end{align}
and similarly, once $T_0$, $T_1$ and $T_2$ are known, the next two Chebyshev polynomials can be computed via:
\begin{align}
\begin{split}\label{eq:cheby2}
    T_3 & = 2T_2T_1 - T_1\\
    T_4 & = 2T_2T_2 - T_0\;,
\end{split}
\end{align}
using only previously known data from \cref{eq:cheby1}. Each Chebyshev polynomial on the left in \cref{eq:cheby2} can therefore be computed in parallel. In general, the low-order Chebyshev expansion $S_j$ of length $k$ will require $\lceil \log_2(k) \rceil$ serial stages to calculate all the required $T_i$'s, where the polynomials within each stage use only already known data so that they can be computed in parallel. To implement this parallelism on GPUs, we use multiple CUDA or HIP streams, for Nvidia and AMD GPUs respectively, as accessible through the MAGMA interface (using the ${\tt magma\_queue}$ struct). An algorithm which computes the Chebyshev polynomials $T_i$, $i=2,\dots, k$ is described using pseudocode in \cref{algorithm:T_i}. 

\begin{figure}
    \centering
    \begin{tikzpicture}
        \begin{axis}[width=.485\textwidth,height=2.0in,
                    xmin=5, xmax=1050,
                    ymin=1.0, ymax=2.0,
                    legend pos=north west,
                    legend columns=3,
                    ylabel={Speed up factor},
                    xlabel={Number of Chebyshev expansion terms $L$},
                    ymajorgrids=true,
                    grid style=dashed,
                    legend entries={$N=500$, $N=600$, $N=700$,$N=800$, $N=900$, $N=1000$}]
        \addplot[blue,mark=square*,line width=0.5] table {data/500-v100.dat};
        \addplot[red,mark=triangle*, line width=0.5] table {data/600-v100.dat};
        \addplot[black,mark=diamond*, line width=0.5] table {data/700-v100.dat};
        \addplot[orange,mark=otimes*, line width=0.5] table {data/800-v100.dat};
        \addplot[purple,mark=*, line width=0.5] table {data/900-v100.dat};
        \addplot[brown,mark=*, line width=0.5] table {data/1000-v100.dat};
        \end{axis}
    \end{tikzpicture}\\
    \begin{tikzpicture}
        \begin{axis}[width=.485\textwidth,height=2.0in,
                    xmin=5, xmax=1050,
                    ymin=1.0, ymax=2.0,
                    legend pos=north west,
                    legend columns=3,
                    ylabel={Speed up factor},
                    xlabel={Number of Chebyshev expansion terms $L$},
                    ymajorgrids=true,
                    grid style=dashed,
                    legend entries={$N=500$, $N=600$, $N=700$,$N=800$, $N=900$, $N=1000$}]
        \addplot[blue,mark=square*,line width=0.5] table {data/crusher/500.dat};
        \addplot[red,mark=triangle*, line width=0.5] table {data/crusher/600.dat};
        \addplot[black,mark=diamond*, line width=0.5] table {data/crusher/700.dat};
        \addplot[orange,mark=otimes*, line width=0.5] table {data/crusher/800.dat};
        \addplot[purple,mark=*, line width=0.5] table {data/crusher/900.dat};
        \addplot[brown,mark=*, line width=0.5] table {data/crusher/1000.dat};
        \end{axis}
    \end{tikzpicture}
        \begin{tikzpicture}
        \begin{semilogyaxis}[width=.485\textwidth,height=2.5in,
                    xmin=5, xmax=1050,
                    ymin=5e-4, ymax=5,
                    legend pos=south east,
                    legend columns=2,
                    ylabel={Time to solution (s)},
                    xlabel={Number of Chebyshev expansion terms},
                    ymajorgrids=true,
                    grid style=dashed,
                    domain=100:1000,
                    legend entries={$N=500$, 
                    $N=1000$,
                    $N=2000$, 
                    $N=4000$}]
        \addplot[blue,line width=0.5] table {data/v100_solution_time-500.dat};
        \addplot[brown,line width=0.75] table 
        {data/v100_solution_time-1000.dat};  
        \addplot[red, line width=1.0] table {data/v100_solution_time-2000.dat};
        \addplot[black, line width=0.75] table {data/v100_solution_time-4000.dat};
        \addplot[blue,line width=1.0,dashed] table {data/crusher/amd_solution_time-500.dat};
        \addplot[brown,line width=1.0,dashed] table 
        {data/crusher/amd_solution_time-1000.dat};  
        \addplot[red, line width=1.0,dashed] table {data/crusher/amd_solution_time-2000.dat};
        \addplot[black,line width=1.0,dashed] table {data/crusher/amd_solution_time-4000.dat};
        \end{semilogyaxis}
    \end{tikzpicture}
    \caption{(top) Speed up of parallelized Chebyshev expansion for matrices of size $N \times N$ when using multiple GPU streams versus using only a single GPU stream. In these tests, $k=m=\sqrt{L+1}$. Calculations were run on Summit's Nvidia V100 Power9 nodes using a single GPU. 
    middle) Same as the top plot but for an AMD MI250X GPU on Crusher. (bottom) Time-to-solution for construction of the single-particle density matrix as a function of expansion size on an Nvidia V100 (solid lines) and AMD MI250X (dashed lines) using multiple streams for $N=500$ and $N=1000$. Only a single stream is used for $N=2000$ and $N=4000$. 
    } \label{fig:speed up_parallel}
\end{figure}
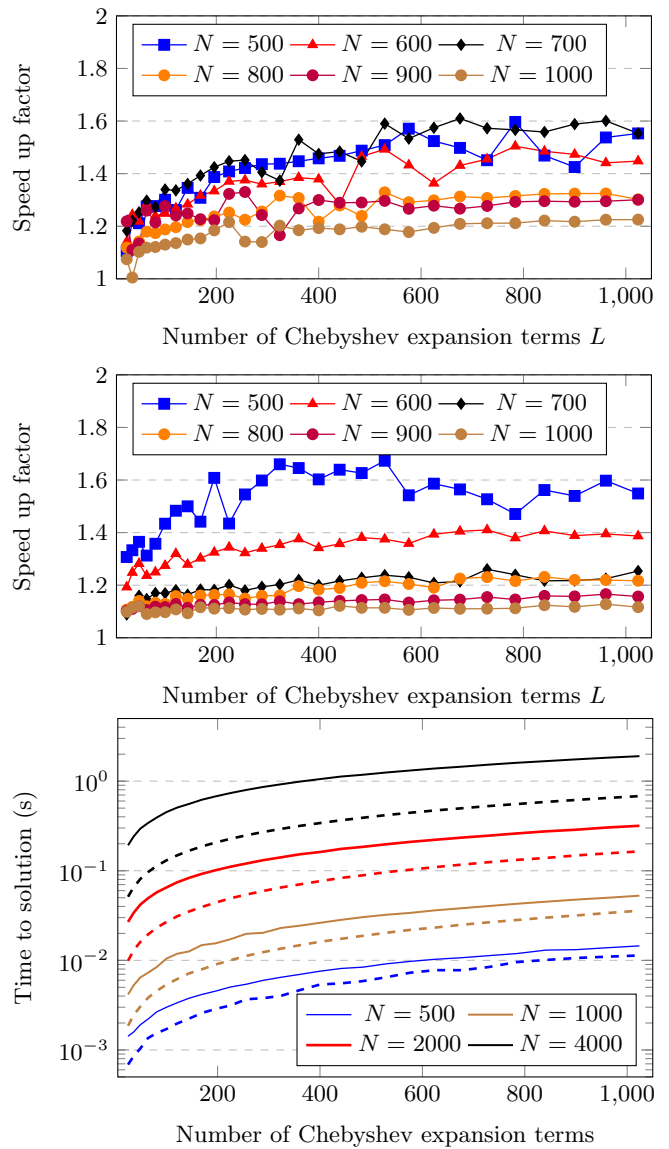

With these two parallel adaptations, we improve on the gain in speed over diagonalization for smaller matrix sizes. The top portion of \cref{fig:speed up_parallel} shows the speed up in building a Chebyshev expansion approximation to the density matrix that is due solely to parallelization and the implementation using GPU streams. In our set of examples, the most prominent effect for the V100 is for matrices of size $700 \times 700$, as we gain approximately 1.6x in speed for polynomial expansions of over one thousand terms. Our GPU stream implementation becomes less efficient as the matrix sizes are increased beyond that. At $1000 \times 1000$, a 1,024 term expansion experiences a more modest 1.2x speed up. For much larger matrix sizes, such as $2000 \times 2000$ (not shown), the benefit from GPU streams is entirely negligible. It is expected that the concurrent computation of several matrix multiplications is not going to give much speed up when a single matrix multiplication is large enough to fully utilize GPU resources. We also expect that the matrix sizes where streams are efficient versus inefficient will vary based on the GPU architecture being used. Indeed, for the MI250X, HIP streams exhibit slightly different behavior. The largest speed up now occurs for $500 \times 500$ matrices, yielding close to 1.6x speed up for large expansions. The speed ups then seem to monotonically decrease as the matrix size is increased. With a $1000 \times 1000$ matrix, a 1,024 term expansion only yields a 1.15x speed up with multiple streams, less than what can be gotten on a V100.

Finally, \cref{fig:speed up_parallel} (bottom) shows the time-to-solution for various matrix sizes and expansion orders for the $\mathcal{O} (\sqrt{L})$ algorithm using GPU streams on both Nvidia V100 and AMD MI250X GPUs. The algorithm is able to take advantage of the highest flops rate of the AMD architecture for the largest matrix sizes ($N$=2000 and $N$=4000), while the difference between the two GPUs is less significant for smaller matrix sizes. All times are close to or below one second, even for the largest matrices ($N$=4000) and the longest polynomial expansion ($L$=1024). When examining our timings, we also observe that after reducing the number of matrix-matrix multiplications with the $\mathcal{O}(\sqrt{L})$ expansion method, our solver spends a significant portion of the time in matrix additions. This is also due to the fact that matrix additions have a less favorable flops to memory access ratio, or arithmetic intensity, which makes them almost as costly as matrix multiplications.  

\begin{algorithm}
\For{$1 \le s \le $ number of stages}{
\If{$s=1$}{
    $T_2 = 2T_1T_1 - T_0$, on stream $0$;
}
    $u=2^{s-2}, v=2^{s-1}$ \\
\For{$u < i \le v $}{
\If{$i+u \le k$}{
    $T_{i+u} = 2T_iT_u-T_{i-u}$, on stream $i-u-1$;
}
}
\For{$u < i \le v $}{
\If{$i+v \le k$}{
     $T_{i+v} = 2T_vT_i-T_{v-i}$, on stream $i-1$;
}
}
\tcp{wait for all streams before next stage}
sync\underline{\hspace{0.25cm}}streams()
}
\caption{The algorithm to compute the Chebyshev polynomials in the low-order expansions of size $k$, $S_j$, using GPU concurrency (CUDA/HIP streams).}\label{algorithm:T_i}
\end{algorithm}

\subsection{Model tight-binding Hamiltonian matrices}
\label{sec:results}

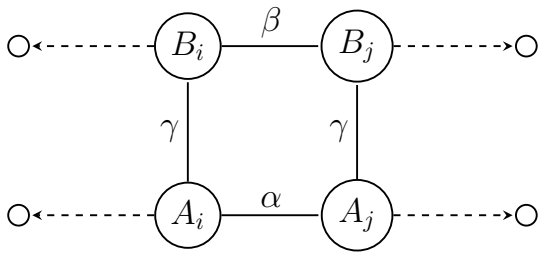
\begin{figure}[h]
    \centering
\scalebox{.9}{
    \begin{tikzpicture}[->,>=stealth,shorten >=1pt,auto,node distance=2.5cm,
                thick,main node/.style={circle,draw,font=\Large}]
  \node[main node] (b) {$B_i$};
  \node[main node] (a) [below of=b] {$A_i$};
  \node[main node] (b2) [right of=b] {$B_j$};
   \node[main node] (a2) [below of=b2] {$A_j$};
   \node[main node] (p1) [left of=b] {};
   \node[main node] (p2) [right of=b2] {};
   \node[main node] (p3) [left of=a] {};
   \node[main node] (p4) [right of=a2] {};

  \path[-]
    (a) edge node [dashed] {\Large $\gamma$} (b)
    (a2) edge node {\Large $\gamma$} (b2)
    (b) edge node {\Large $\beta$} (b2)
    (a) edge node {\Large $\alpha$} (a2);
  \path[dashed]
    (b) edge (p1)
    (b2) edge (p2)
    (a) edge (p3)
    (a2) edge (p4)
    ;
\end{tikzpicture}
}
    \caption{Schematic representation of the two-level system model used to generate benchmark Hamiltonian matrices. The model includes four coupling parameters, four onsite energies, a decaying exponential parameter, and a randomization factor.}
    \label{fig:modelH}
\end{figure}{}

Hamiltonian matrices that describe chemically relevant systems can generally be classified into one of three categories: metals, semiconductors, and soft matter. Figure~\ref{fig:modelH} provides a schematic representation of the model utilized to generate benchmark Hamiltonian matrices for these three distinct systems.

The construction of these model systems involves coupling two-level systems with atomic-type orbitals {$A$} and {$B$}. The {$A$} and {$B$} orbitals have onsite energies $\epsilon_{{A}}$ and $\epsilon_{{B}}$, respectively. A coupling between orbitals of the same type ({$A$}-{$A$} or {$B$}-{$B$}) is described by $\alpha$ or $\beta$, respectively.
Likewise, a coupling between two different orbital types, {$A$} and {$B$}, is denoted by $\gamma$. Additionally, all couplings between orbitals are subject to modulation by an exponential decay factor $\exp(-k|i-j|)$, where $k$ is a decay constant, and $i$ and $j$ represent the matrix positions of the two respective orbitals. In the case of soft matter type Hamiltonians, we introduce a randomization parameter that adds noise to both couplings and onsite energies. Specifically, we apply a multiplicative coefficient of $(1 + r \times \eta)$, where $\eta$ is a uniformly distributed random number between -1 and 1 and $r$ is an adjustable amplitude parameter. A module in the PROGRESS~\cite{2022progress} library facilitates the generation of these model Hamiltonian matrices. 

\begin{figure}
    \vspace{1cm}
    \includegraphics[scale=0.3]{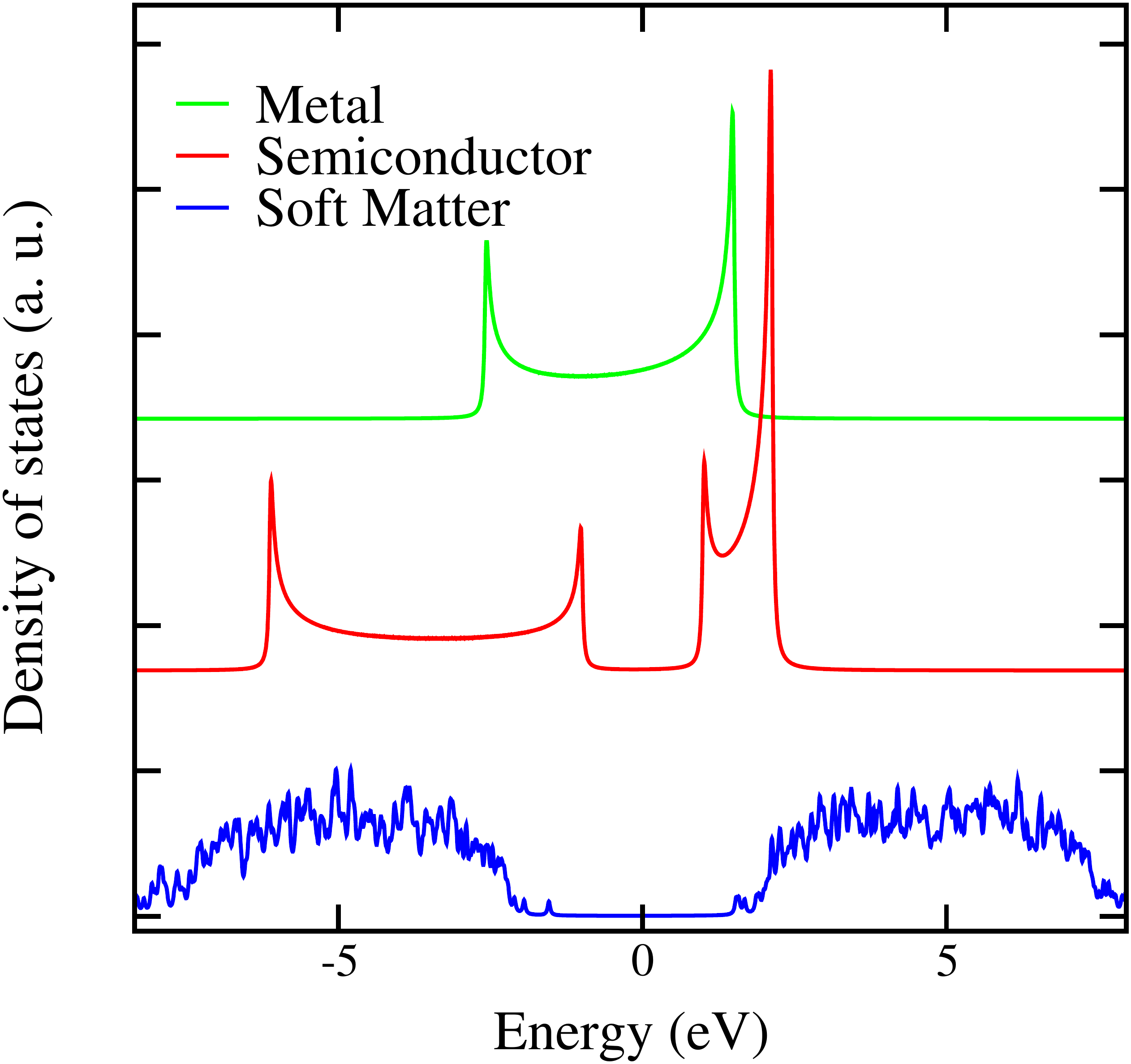}
    \caption{Total density of states (in arbitrary units) computed with different model Hamiltonian matrices representing metals (green), semiconductors (red), and soft matter (blue) systems. Plots were shifted along the $y$-axis to facilitate comparison. The Fermi level of the system is set to be 0.0 eV. Model Hamiltonian parameter values are given in the text.}
    \label{fig:systemsH}
\end{figure}{}

Examples on how to set the parameters to obtain Hamiltonian matrices representing different systems are given below.
To generate system matrices representing metals, we can set the parameters as follows: $\alpha = -1.0$, $\beta = -1.0$, and $k = -1.0$, while the remaining parameters set to 0.0. For semiconductors, we set the parameters to: $\beta = -1.0$,  $\gamma = -2.0$, and $k = -0.01$, with the remaining parameters set to 0.0. Lastly, for soft matter system, matrices can be generated using: $\beta = -1.0$, $\gamma = -1.0$, $\epsilon_A = -10.0$,  $k = -0.1$, and $r = 1.0$, and all remaining parameters set to 0.0. Figure~\ref{fig:systemsH} displays a plot of the resulting total density of states (DOS) for the three types of model Hamiltonian matrices, where all couplings and onsite energies have units of electron volts (eV). In each case, the Fermi level of the system is set to 0.0 eV.

\subsection{Accuracy}

In order to test both accuracy and numerical stability of our algorithm, we use a model Hamiltonian as just described with parameters chosen in order to obtain characteristic metallic behavior --- a DOS concentrated around the chemical potential. Parameters $\alpha$, $\beta$, $\gamma$ and $k$, were set to $-1.0$, $1.0$, $0.0$, and $-1.0$ respectively. The onsite energies were set to $\epsilon_a=1.0$ and $\epsilon_b=-1.0$. This type of Hamiltonian was specifically picked to be challenging since it has a wide spectral range and the DOS is heavily concentrated around the Fermi level. Systems with a large number of states near the Fermi level are known to be difficult to solve with spectral purification methods. A recent report \cite{MOHR201864} on electronic structure calculations using Chebyshev polynomial expansions applied to metallic systems suggested using polynomial orders in-between 400 and 1300.

\begin{figure}
    \centering
    \begin{tikzpicture}
        \begin{semilogyaxis}[width=.485\textwidth,height=2.5in,
                    xmin=10, xmax=512,
                    ymin=1e-8, ymax=1.0,
                    legend pos=north east,
                    legend columns=1,
                    ylabel={Relative accuracy},
                    xlabel={Number of Chebyshev expansion terms},
                    ymajorgrids=true,
                    grid style=dashed,
                    legend entries={$800 \times 800$}]
        \addplot[blue,mark=square*,line width=0.25] table {data/accuracy.dat};
        \end{semilogyaxis}
    \end{tikzpicture}
    \caption{Relative error, in the Frobenius norm, of the density matrix approximation at $k_bT = 0.1$ eV using a metallic-like Hamiltonian and the $\mathcal{O}(\sqrt{L})$ Chebyshev expansion. In this example, the spectral range is $\varepsilon_b-\varepsilon_a \approx 103$ eV. A system size of $800 \times 800$ is used and the reference density matrix is calculated with diagonalization.}
    \label{fig:accuracy}
\end{figure}
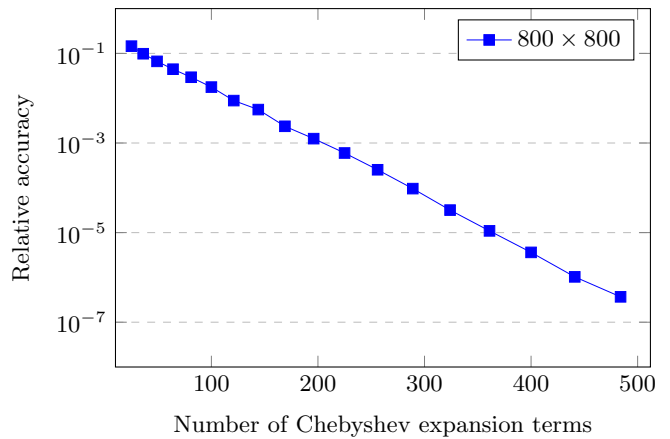

In \cref{fig:accuracy} we show the relative accuracy as a function of the order of the polynomial expansion for an 800 $\times$ 800 Hamiltonian matrix. The accuracy was computed using the relative Frobenius norm of the difference between the exact density matrix obtained with diagonalization and an approximation using a Chebyshev expansion. \Cref{fig:accuracy} shows an exponential error decay as a function of the number of terms. The combination of a Chebyshev expansion's Gibb's oscillations and the high number of states near the chemical potential results in the need to use many terms in the expansion. With approximately 500 expansion terms, the relative error is around $10^{-7}$. This result indicates that, for small metallic-like systems (where direct dense diagonalization on GPU is not efficient) the $\mathcal{O}(\sqrt{L})$ Chebyshev algorithm performs well in terms of accuracy and stability. Based on the speed up results presented in \cref{fig:speed up_diag,fig:speed up_parallel}, we would roughly expect a $\sim$40x speed up over a cuSolver diagonalization for this system on a V100 GPU.

\section{Conclusion}
We have implemented and investigated a Chebyshev polynomial expansion which minimizes the number of matrix multiplications, as an alternative to computing the density matrix used in electronic structure calculations on GPUs. Our findings shows that the unique combination of the $\mathcal{O}(\sqrt{L})$ matrix multiplication algorithm, together with concurrency tools present on GPUs, lead to significant speed ups compared to using a dense diagonalization. We have also shown, that, even for cases where the electronic structure is difficult to compute (i.e. metallic-like systems) this $\mathcal{O}(\sqrt{L})$ Chebyshev expansion is numerically stable and has a well controlled accuracy as a function of the degree of the polynomial expansion. This technique can be readily implemented into localized atomic orbital-based Quantum Chemistry packages such as DFTB+ \cite{dftb+}, LATTE \cite{latte} and Siesta \cite{Soler2002-wc}.

\section*{Acknowledgements}

This research used resources of the Oak Ridge Leadership Computing Facility at the Oak Ridge National Laboratory, which is supported by the Office of Science of the U.S. Department of Energy under Contract No. DE-AC05-00OR22725. This research was supported by the Exascale Computing Project (17-SC-20-SC), a collaborative effort of the DOE Office of Science and the National Nuclear Security Administration (NNSA). The authors thank Anders M. Niklasson for his helpful comments and suggestions on the manuscript.

\section*{Author declarations}
The authors have no conflict of interest to disclose.

\section*{Data availability}
The data and code supporting the findings of this study are available from the corresponding authors upon reasonable request.

\section*{Appendix: Derivation of upper-triangular block matrix}

\begin{widetext}

To solve  \cref{eq:formula for d0} and \cref{eq:formula for d}, we organize the equations into a larger $km\times km$ matrix-vector equation. This allows us to solve for all $d_{i,j}$ at once. For each $0\le \ell \le m-1$ and $0 < i \le k-1$, one has that
\begin{align}\label{eq:dot_prod}
    {\bf d}_i^T {\bf a}_\ell & = a _{0,\ell} d_{i,0} +  a_{1,\ell} d_{i,1}  + \cdots +  a_{m-1,\ell} d_{i,m-1}  \;.
\end{align}
Using the fact that ${\bf A}$ is lower-triangular, when $\ell=m-1$, we have
\begin{align*}
    {\bf d}_0^T {\bf a}_{m-1} & =  a_{m-1,m-1} d_{0,m-1}\\
    {\bf d}_1^T {\bf a}_{m-1} & =  a_{m-1,m-1} d_{1,m-1}\\
    {\bf d}_2^T {\bf a}_{m-1} & =  a_{m-1,m-1} d_{2,m-1}\\
    \vdots \\
    {\bf d}_{k-1} ^T {\bf a}_{m-1} & =  a_{m-1,m-1} d_{k-1,m-1}\\
\end{align*}
so that,
\begin{align*}
    \begin{bmatrix}
        c_{(m-1) \times k}\\
        c_{(m-1) \times k+1}\\
        \vdots\\
        c_{(m-1) \times k + (k-1)}
    \end{bmatrix} 
    =\begin{bmatrix}
        a_{m-1,m-1}  &                                  \\
                     & \tfrac{1}{2} a_{m-1,m-1} {\bf I} \\
    \end{bmatrix}
    \begin{bmatrix}
        d_{0,m-1} \\
        d_{1,m-1} \\
        \vdots \\
        d_{k-1,m-1} \\
    \end{bmatrix} 
\end{align*}
for ${\bf I}$ the $k-1\times k-1$ identity matrix. Next, for $\ell=m-2$, we use the equation in \cref{eq:dot_prod} for each $i$ to get
\begin{align*}
    {\bf d}_0^T {\bf a}_{m-2}  & =  a_{m-2,m-2} d_{0,m-2} + a_{m-1,m-2} d_{0,m-1}  \\
    & =  a_{m-2,m-2} d_{0,m-2}   \\
    {\bf d}_1^T {\bf a}_{m-2} + {\bf d}_{k-1}^T {\bf a}_{m-1} & =  a_{m-2,m-2} d_{1,m-2} + a_{m-1,m-2} d_{1,m-1} + a_{m-1,m-1} d_{k-1,m-1} \\
    &= a_{m-2,m-2} d_{1,m-2} + a_{m-1,m-1} d_{k-1,m-1}\\
    {\bf d}_2^T {\bf a}_{m-2} + {\bf d}_{k-2}^T {\bf a}_{m-1} & =  a_{m-2,m-2} d_{2,m-2} + a_{m-1,m-2} d_{2,m-1} + a_{m-1,m-1} d_{k-2,m-1} \\
    &= a_{m-2,m-2} d_{2,m-2} + a_{m-1,m-1} d_{k-2,m-1} \\
    \vdots \hspace{-2.5in} \vdots \\
    {\bf d}_{k-1}^T {\bf a}_{m-2} + {\bf d}_{1}^T {\bf a}_{m-1} & =  a_{m-2,m-2} d_{k-1,m-2} + a_{m-1,m-2} d_{k-1,m-1} + a_{m-1,m-1} d_{1,m-2} \\
    & = a_{m-2,m-2} d_{k-1,m-2}  + a_{m-1,m-1} d_{1,m-1} \\
\end{align*}
because $a_{m-1,m-2}=0$ by construction of ${\bf A}$ (see the example in \cref{eq:example A}). Therefore,
\begin{align*}
    \begin{bmatrix}
        c_{(m-2) \times k}\\
        c_{(m-2) \times k+1}\\
        \vdots\\
        c_{(m-2) \times k + (k-1)}\\
        c_{(m-1) \times k}\\
        c_{(m-1) \times k+1}\\
        \vdots\\
        c_{(m-1) \times k + (k-1)}
    \end{bmatrix} 
    = 
    \begin{bmatrix}
    a_{m-2,m-2} &  & 0 &  \\
    & \tfrac{1}{2} a_{m-2,m-2} {\bf I} &  & \tfrac{1}{2} a_{m-1,m-1} {\bf J} \\
    & & a_{m-1,m-1}  &  \\
    & & & \tfrac{1}{2} a_{m-1,m-1} {\bf I} 
    \end{bmatrix}
    \begin{bmatrix}
        d_{0,m-2} \\
        d_{1,m-2} \\
        \vdots \\
        d_{k-1,m-2} \\
        d_{0,m-1} \\
        d_{1,m-1} \\
        \vdots \\
        d_{k-1,m-1} \\
    \end{bmatrix} 
\end{align*}
where only the upper triangular part of the matrix is explicitly shown (the rest is zero), and ${\bf J}$ is the identity matrix with the columns reversed, i.e.
\begin{align*}
    {\bf J}=
    \begin{pmatrix}
        &   &         &   &  1  \\
        &   &         & 1 &     \\
        &   &  \iddots &   &     \\ 
        & 1 &         &   &     \\ 
     1  &   &         &   &     \\ 
    \end{pmatrix}\;.
\end{align*}
Continuing to the next case of $\ell = m-3$, and using the fact that $a_{m-2,m-3} = 0$ (and in general that $a_{i,j} = 0$ if $i$ and $j$ have different parity), we come to
\begin{align*}
    \begin{bmatrix}
        c_{(m-3) \times k}\\
        c_{(m-3) \times k+1}\\
        \vdots\\
        c_{(m-3) \times k + (k-1)}\\
        c_{(m-2) \times k}\\
        c_{(m-2) \times k+1}\\
        \vdots\\
        c_{(m-2) \times k + (k-1)}\\
        c_{(m-1) \times k}\\
        c_{(m-1) \times k+1}\\
        \vdots\\
        c_{(m-1) \times k + (k-1)}
    \end{bmatrix} 
    = 
    \begin{bmatrix}
    a_{m-3,m-3} &  & 0 &  & a_{m-1,m-3} & \\
    & \tfrac{1}{2} a_{m-3,m-3} {\bf I} &  & \tfrac{1}{2} a_{m-2,m-2} {\bf J}  &  & \tfrac{1}{2} a_{m-1,m-3} {\bf I} \\
    & &a_{m-2,m-2} &  & 0 &  \\
    & & & \tfrac{1}{2} a_{m-2,m-2} {\bf I} &  & \tfrac{1}{2} a_{m-1,m-1} {\bf J} \\
    & & & & a_{m-1,m-1}  &  \\
    & & & & & \tfrac{1}{2} a_{m-1,m-1} {\bf I} 
    \end{bmatrix}
    \begin{bmatrix}
        d_{0,m-3} \\
        d_{1,m-3} \\
        \vdots \\
        d_{k-1,m-3} \\
        d_{0,m-2} \\
        d_{1,m-2} \\
        \vdots \\
        d_{k-1,m-2} \\
        d_{0,m-1} \\
        d_{1,m-1} \\
        \vdots \\
        d_{k-1,m-1} \\
    \end{bmatrix} \;.
\end{align*}
To complete the derivation, this process is continued until we reach the case $\ell = m-m=0$ where the dot 
products in \cref{eq:dot_prod} are (assuming $m$ is odd, the even case is similar),
\begin{align*}
    {\bf d}_0^T {\bf a}_{0}  & =  a_{0,0} d_{0,0} + a_{2,0} d_{0,2}  + \cdots + a_{m-3,0} d_{0,m-3} + a_{m-1,0} d_{0,m-1}\\
    {\bf d}_1^T {\bf a}_{0} + \tfrac{1}{2}{\bf d}_{k-1}^T {\bf a}_{1} & =  (a_{0,0} d_{1,0} + a_{2,0} d_{1,2} + \cdots + a_{m-1,0} d_{1,m-1}) + \tfrac{1}{2} (a_{1,1} d_{k-1,1} + a_{3,1} d_{k-1,3} + \cdots + a_{m-2,1} d_{k-1,m-2})\\
    {\bf d}_2^T {\bf a}_{0} + \tfrac{1}{2}{\bf d}_{k-2}^T {\bf a}_{1} & =  (a_{0,0} d_{2,0} + a_{2,0} d_{2,2} + \cdots + a_{m-1,0} d_{2,m-1}) + \tfrac{1}{2} (a_{1,1} d_{k-2,1} + a_{3,1} d_{k-2,3} + \cdots + a_{m-2,1} d_{k-2,m-2})\\
    \vdots \hspace{-2.5in} \vdots \\
    {\bf d}_{k-1}^T {\bf a}_{0} + \tfrac{1}{2}{\bf d}_{1}^T {\bf a}_{1} & =   (a_{0,0} d_{k-1,0} + a_{2,0} d_{k-1,2} + \cdots + a_{m-1,0} d_{k-1,m-1}) \\ 
    & \hspace{7cm}+ \tfrac{1}{2} (a_{1,1} d_{1,1} + a_{3,1} d_{1,3} + \cdots + a_{m-2,1} d_{1,m-2})\\
\end{align*}
so that the first row block in ${\bf X}$ of \cref{eq:X} is 
\begin{align}
    \begin{bmatrix}
        a_{0,0}  & & 0 & & a_{2,0} & & 0 & & a_{4,0} & \cdots & 0& & &  a_{m-1,0} &  \\
                 & a_{0,0} {\bf I} & & \tfrac{1}{2} a_{1,1} {\bf J} & & a_{2,0} {\bf I}& & \tfrac{1}{2} a_{3,1} {\bf J}&  & \dots & & & \tfrac{1}{2}a_{m-2,1} {\bf J} & & a_{m-1,0} {\bf I}
    \end{bmatrix} \;.
\end{align}

\end{widetext}

\section*{References}
\bibliography{bib}

\providecommand{\noopsort}[1]{}\providecommand{\singleletter}[1]{#1}%
\begin{thebibliography}{32}%
\makeatletter
\providecommand \@ifxundefined [1]{%
 \@ifx{#1\undefined}
}%
\providecommand \@ifnum [1]{%
 \ifnum #1\expandafter \@firstoftwo
 \else \expandafter \@secondoftwo
 \fi
}%
\providecommand \@ifx [1]{%
 \ifx #1\expandafter \@firstoftwo
 \else \expandafter \@secondoftwo
 \fi
}%
\providecommand \natexlab [1]{#1}%
\providecommand \enquote  [1]{``#1''}%
\providecommand \bibnamefont  [1]{#1}%
\providecommand \bibfnamefont [1]{#1}%
\providecommand \citenamefont [1]{#1}%
\providecommand \href@noop [0]{\@secondoftwo}%
\providecommand \href [0]{\begingroup \@sanitize@url \@href}%
\providecommand \@href[1]{\@@startlink{#1}\@@href}%
\providecommand \@@href[1]{\endgroup#1\@@endlink}%
\providecommand \@sanitize@url [0]{\catcode `\\12\catcode `\$12\catcode
  `\&12\catcode `\#12\catcode `\^12\catcode `\_12\catcode `\%12\relax}%
\providecommand \@@startlink[1]{}%
\providecommand \@@endlink[0]{}%
\providecommand \url  [0]{\begingroup\@sanitize@url \@url }%
\providecommand \@url [1]{\endgroup\@href {#1}{\urlprefix }}%
\providecommand \urlprefix  [0]{URL }%
\providecommand \Eprint [0]{\href }%
\providecommand \doibase [0]{https://doi.org/}%
\providecommand \selectlanguage [0]{\@gobble}%
\providecommand \bibinfo  [0]{\@secondoftwo}%
\providecommand \bibfield  [0]{\@secondoftwo}%
\providecommand \translation [1]{[#1]}%
\providecommand \BibitemOpen [0]{}%
\providecommand \bibitemStop [0]{}%
\providecommand \bibitemNoStop [0]{.\EOS\space}%
\providecommand \EOS [0]{\spacefactor3000\relax}%
\providecommand \BibitemShut  [1]{\csname bibitem#1\endcsname}%
\let\auto@bib@innerbib\@empty
\bibitem [{\citenamefont {Niklasson}(2002)}]{ANiklasson02}%
  \BibitemOpen
  \bibfield  {author} {\bibinfo {author} {\bibfnamefont {A.~M.}\ \bibnamefont
  {Niklasson}},\ }\bibfield  {title} {\enquote {\bibinfo {title} {Expansion
  algorithm for the density matrix},}\ }\href@noop {} {\bibfield  {journal}
  {\bibinfo  {journal} {Phys. Rev. B}\ }\textbf {\bibinfo {volume} {66}},\
  \bibinfo {pages} {155115} (\bibinfo {year} {2002})}\BibitemShut {NoStop}%
\bibitem [{\citenamefont {Niklasson}(2003)}]{ANiklasson03}%
  \BibitemOpen
  \bibfield  {author} {\bibinfo {author} {\bibfnamefont {A.~M.~N.}\
  \bibnamefont {Niklasson}},\ }\bibfield  {title} {\enquote {\bibinfo {title}
  {Implicit purification for temperature-dependent density matrices},}\ }\href
  {https://doi.org/10.1103/PhysRevB.68.233104} {\bibfield  {journal} {\bibinfo
  {journal} {Phys. Rev. B}\ }\textbf {\bibinfo {volume} {68}},\ \bibinfo
  {pages} {233104} (\bibinfo {year} {2003})}\BibitemShut {NoStop}%
\bibitem [{\citenamefont {Goedecker}\ and\ \citenamefont
  {Teter}(1995)}]{GoedeckerTeter1995}%
  \BibitemOpen
  \bibfield  {author} {\bibinfo {author} {\bibfnamefont {S.}~\bibnamefont
  {Goedecker}}\ and\ \bibinfo {author} {\bibfnamefont {M.}~\bibnamefont
  {Teter}},\ }\bibfield  {title} {\enquote {\bibinfo {title} {Tight-binding
  electronic-structure calculations and tight-binding molecular dynamics with
  localized orbitals},}\ }\href {https://doi.org/10.1103/PhysRevB.51.9455}
  {\bibfield  {journal} {\bibinfo  {journal} {Phys. Rev. B}\ }\textbf {\bibinfo
  {volume} {51}},\ \bibinfo {pages} {9455--9464} (\bibinfo {year}
  {1995})}\BibitemShut {NoStop}%
\bibitem [{\citenamefont {Baer}\ and\ \citenamefont
  {Head-Gordon}(1997)}]{Baer1997}%
  \BibitemOpen
  \bibfield  {author} {\bibinfo {author} {\bibfnamefont {R.}~\bibnamefont
  {Baer}}\ and\ \bibinfo {author} {\bibfnamefont {M.}~\bibnamefont
  {Head-Gordon}},\ }\bibfield  {title} {\enquote {\bibinfo {title} {Sparsity of
  the density matrix in {Kohn-Sham} density functional theory and an assessment
  of linear system-size scaling methods},}\ }\href
  {https://doi.org/10.1103/PhysRevLett.79.3962} {\bibfield  {journal} {\bibinfo
   {journal} {Phys. Rev. Lett.}\ }\textbf {\bibinfo {volume} {79}},\ \bibinfo
  {pages} {3962--3965} (\bibinfo {year} {1997})}\BibitemShut {NoStop}%
\bibitem [{\citenamefont {Aarons}\ and\ \citenamefont
  {Skylaris}(2018)}]{AaronsSkylaris2018}%
  \BibitemOpen
  \bibfield  {author} {\bibinfo {author} {\bibfnamefont {J.}~\bibnamefont
  {Aarons}}\ and\ \bibinfo {author} {\bibfnamefont {C.-K.}\ \bibnamefont
  {Skylaris}},\ }\bibfield  {title} {\enquote {\bibinfo {title} {Electronic
  annealing {Fermi} operator expansion for {DFT} calculations on metallic
  systems},}\ }\href {https://doi.org/10.1063/1.5001340} {\bibfield  {journal}
  {\bibinfo  {journal} {J. Chem. Phys.}\ }\textbf {\bibinfo {volume} {148}},\
  \bibinfo {pages} {074107} (\bibinfo {year} {2018})}\BibitemShut {NoStop}%
\bibitem [{\citenamefont {Mohr}\ \emph {et~al.}(2018)\citenamefont {Mohr},
  \citenamefont {Eixarch}, \citenamefont {Amsler}, \citenamefont {Mantsinen},\
  and\ \citenamefont {Genovese}}]{MOHR201864}%
  \BibitemOpen
  \bibfield  {author} {\bibinfo {author} {\bibfnamefont {S.}~\bibnamefont
  {Mohr}}, \bibinfo {author} {\bibfnamefont {M.}~\bibnamefont {Eixarch}},
  \bibinfo {author} {\bibfnamefont {M.}~\bibnamefont {Amsler}}, \bibinfo
  {author} {\bibfnamefont {M.~J.}\ \bibnamefont {Mantsinen}},\ and\ \bibinfo
  {author} {\bibfnamefont {L.}~\bibnamefont {Genovese}},\ }\bibfield  {title}
  {\enquote {\bibinfo {title} {Linear scaling {DFT} calculations for large
  tungsten systems using an optimized local basis},}\ }\href
  {https://doi.org/https://doi.org/10.1016/j.nme.2018.01.002} {\bibfield
  {journal} {\bibinfo  {journal} {Nuclear Materials and Energy}\ }\textbf
  {\bibinfo {volume} {15}},\ \bibinfo {pages} {64--70} (\bibinfo {year}
  {2018})}\BibitemShut {NoStop}%
\bibitem [{\citenamefont {Zhou}\ \emph {et~al.}(2006)\citenamefont {Zhou},
  \citenamefont {Saad}, \citenamefont {Tiago},\ and\ \citenamefont
  {Chelikowsky}}]{Zhou2006}%
  \BibitemOpen
  \bibfield  {author} {\bibinfo {author} {\bibfnamefont {Y.}~\bibnamefont
  {Zhou}}, \bibinfo {author} {\bibfnamefont {Y.}~\bibnamefont {Saad}}, \bibinfo
  {author} {\bibfnamefont {M.~L.}\ \bibnamefont {Tiago}},\ and\ \bibinfo
  {author} {\bibfnamefont {J.~R.}\ \bibnamefont {Chelikowsky}},\ }\bibfield
  {title} {\enquote {\bibinfo {title} {Parallel self-consistent-field
  calculations via {Chebyshev}-filtered subspace acceleration},}\ }\href
  {https://doi.org/10.1103/PhysRevE.74.066704} {\bibfield  {journal} {\bibinfo
  {journal} {Phys. Rev. E}\ }\textbf {\bibinfo {volume} {74}},\ \bibinfo
  {pages} {066704} (\bibinfo {year} {2006})}\BibitemShut {NoStop}%
\bibitem [{\citenamefont {Alemany}\ \emph {et~al.}(2007)\citenamefont
  {Alemany}, \citenamefont {Jain}, \citenamefont {Tiago}, \citenamefont {Zhou},
  \citenamefont {Saad},\ and\ \citenamefont {Chelikowsky}}]{ALEMANY2007}%
  \BibitemOpen
  \bibfield  {author} {\bibinfo {author} {\bibfnamefont {M.}~\bibnamefont
  {Alemany}}, \bibinfo {author} {\bibfnamefont {M.}~\bibnamefont {Jain}},
  \bibinfo {author} {\bibfnamefont {M.~L.}\ \bibnamefont {Tiago}}, \bibinfo
  {author} {\bibfnamefont {Y.}~\bibnamefont {Zhou}}, \bibinfo {author}
  {\bibfnamefont {Y.}~\bibnamefont {Saad}},\ and\ \bibinfo {author}
  {\bibfnamefont {J.~R.}\ \bibnamefont {Chelikowsky}},\ }\bibfield  {title}
  {\enquote {\bibinfo {title} {Efficient first-principles calculations of the
  electronic structure of periodic systems},}\ }\href
  {https://doi.org/https://doi.org/10.1016/j.cpc.2007.04.003} {\bibfield
  {journal} {\bibinfo  {journal} {Comput. Phys. Commun.}\ }\textbf {\bibinfo
  {volume} {177}},\ \bibinfo {pages} {339--347} (\bibinfo {year}
  {2007})}\BibitemShut {NoStop}%
\bibitem [{\citenamefont {White}\ and\ \citenamefont
  {Collins}(2020)}]{AWhite20}%
  \BibitemOpen
  \bibfield  {author} {\bibinfo {author} {\bibfnamefont {A.~J.}\ \bibnamefont
  {White}}\ and\ \bibinfo {author} {\bibfnamefont {L.~A.}\ \bibnamefont
  {Collins}},\ }\bibfield  {title} {\enquote {\bibinfo {title} {Fast and
  universal {Kohn-Sham} density functional theory algorithm for warm dense
  matter to hot dense plasma},}\ }\href@noop {} {\bibfield  {journal} {\bibinfo
   {journal} {Phys. Rev. Lett.}\ }\textbf {\bibinfo {volume} {125}},\ \bibinfo
  {pages} {055002} (\bibinfo {year} {2020})}\BibitemShut {NoStop}%
\bibitem [{\citenamefont {Baer}, \citenamefont {Neuhauser},\ and\ \citenamefont
  {Rabani}(2013)}]{RBaer2013}%
  \BibitemOpen
  \bibfield  {author} {\bibinfo {author} {\bibfnamefont {R.}~\bibnamefont
  {Baer}}, \bibinfo {author} {\bibfnamefont {D.}~\bibnamefont {Neuhauser}},\
  and\ \bibinfo {author} {\bibfnamefont {E.}~\bibnamefont {Rabani}},\
  }\bibfield  {title} {\enquote {\bibinfo {title} {Self-averaging stochastic
  {Kohn-Sham} density-functional theory},}\ }\href@noop {} {\bibfield
  {journal} {\bibinfo  {journal} {Phys. Rev. Lett.}\ }\textbf {\bibinfo
  {volume} {111}},\ \bibinfo {pages} {106402} (\bibinfo {year}
  {2013})}\BibitemShut {NoStop}%
\bibitem [{\citenamefont {Cytter}\ \emph {et~al.}(2018)\citenamefont {Cytter},
  \citenamefont {Rabani}, \citenamefont {Neuhauser},\ and\ \citenamefont
  {Baer}}]{YCytter2018}%
  \BibitemOpen
  \bibfield  {author} {\bibinfo {author} {\bibfnamefont {Y.}~\bibnamefont
  {Cytter}}, \bibinfo {author} {\bibfnamefont {E.}~\bibnamefont {Rabani}},
  \bibinfo {author} {\bibfnamefont {D.}~\bibnamefont {Neuhauser}},\ and\
  \bibinfo {author} {\bibfnamefont {R.}~\bibnamefont {Baer}},\ }\bibfield
  {title} {\enquote {\bibinfo {title} {Stochastic density functional theory at
  finite temperatures},}\ }\href@noop {} {\bibfield  {journal} {\bibinfo
  {journal} {Phys. Rev. B}\ }\textbf {\bibinfo {volume} {97}},\ \bibinfo
  {pages} {115207} (\bibinfo {year} {2018})}\BibitemShut {NoStop}%
\bibitem [{\citenamefont {Sharma}, \citenamefont {Collins},\ and\ \citenamefont
  {White}(2023)}]{VSharma23}%
  \BibitemOpen
  \bibfield  {author} {\bibinfo {author} {\bibfnamefont {V.}~\bibnamefont
  {Sharma}}, \bibinfo {author} {\bibfnamefont {L.~A.}\ \bibnamefont
  {Collins}},\ and\ \bibinfo {author} {\bibfnamefont {A.~J.}\ \bibnamefont
  {White}},\ }\href@noop {} {\enquote {\bibinfo {title} {Stochastic and mixed
  density functional theory within the projector augmented wave formalism for
  the simulation of warm dense matter},}\ } (\bibinfo {year} {2023}),\ \Eprint
  {https://arxiv.org/abs/2301.12018} {arXiv:2301.12018 [physics.comp-ph]}
  \BibitemShut {NoStop}%
\bibitem [{\citenamefont {Nguyen}\ and\ \citenamefont
  {Neuhauser}(2022)}]{MNguyen2022}%
  \BibitemOpen
  \bibfield  {author} {\bibinfo {author} {\bibfnamefont {M.}~\bibnamefont
  {Nguyen}}\ and\ \bibinfo {author} {\bibfnamefont {D.}~\bibnamefont
  {Neuhauser}},\ }\bibfield  {title} {\enquote {\bibinfo {title}
  {Gapped-filtering for efficient {Chebyshev} expansion of the density
  projection operator},}\ }\href
  {https://doi.org/https://doi.org/10.1016/j.cplett.2022.140036} {\bibfield
  {journal} {\bibinfo  {journal} {Chem. Phys. Lett.}\ }\textbf {\bibinfo
  {volume} {806}},\ \bibinfo {pages} {140036} (\bibinfo {year}
  {2022})}\BibitemShut {NoStop}%
\bibitem [{\citenamefont {Mniszewski}\ \emph {et~al.}(2021)\citenamefont
  {Mniszewski}, \citenamefont {Belak}, \citenamefont {Fattebert}, \citenamefont
  {Negre}, \citenamefont {Slattery}, \citenamefont {Adedoyin}, \citenamefont
  {Bird}, \citenamefont {Chang}, \citenamefont {Chen}, \citenamefont {Ethier},
  \citenamefont {Fogerty}, \citenamefont {Habib}, \citenamefont {Junghans},
  \citenamefont {Lebrun-Grandié}, \citenamefont {Mohd-Yusof}, \citenamefont
  {Moore}, \citenamefont {Osei-Kuffuor}, \citenamefont {Plimpton},
  \citenamefont {Pope}, \citenamefont {Reeve}, \citenamefont {Ricketson},
  \citenamefont {Scheinberg}, \citenamefont {Sharma},\ and\ \citenamefont
  {Wall}}]{CoPA2021}%
  \BibitemOpen
  \bibfield  {author} {\bibinfo {author} {\bibfnamefont {S.~M.}\ \bibnamefont
  {Mniszewski}}, \bibinfo {author} {\bibfnamefont {J.}~\bibnamefont {Belak}},
  \bibinfo {author} {\bibfnamefont {J.-L.}\ \bibnamefont {Fattebert}}, \bibinfo
  {author} {\bibfnamefont {C.~F.}\ \bibnamefont {Negre}}, \bibinfo {author}
  {\bibfnamefont {S.~R.}\ \bibnamefont {Slattery}}, \bibinfo {author}
  {\bibfnamefont {A.~A.}\ \bibnamefont {Adedoyin}}, \bibinfo {author}
  {\bibfnamefont {R.~F.}\ \bibnamefont {Bird}}, \bibinfo {author}
  {\bibfnamefont {C.}~\bibnamefont {Chang}}, \bibinfo {author} {\bibfnamefont
  {G.}~\bibnamefont {Chen}}, \bibinfo {author} {\bibfnamefont {S.}~\bibnamefont
  {Ethier}}, \bibinfo {author} {\bibfnamefont {S.}~\bibnamefont {Fogerty}},
  \bibinfo {author} {\bibfnamefont {S.}~\bibnamefont {Habib}}, \bibinfo
  {author} {\bibfnamefont {C.}~\bibnamefont {Junghans}}, \bibinfo {author}
  {\bibfnamefont {D.}~\bibnamefont {Lebrun-Grandié}}, \bibinfo {author}
  {\bibfnamefont {J.}~\bibnamefont {Mohd-Yusof}}, \bibinfo {author}
  {\bibfnamefont {S.~G.}\ \bibnamefont {Moore}}, \bibinfo {author}
  {\bibfnamefont {D.}~\bibnamefont {Osei-Kuffuor}}, \bibinfo {author}
  {\bibfnamefont {S.~J.}\ \bibnamefont {Plimpton}}, \bibinfo {author}
  {\bibfnamefont {A.}~\bibnamefont {Pope}}, \bibinfo {author} {\bibfnamefont
  {S.~T.}\ \bibnamefont {Reeve}}, \bibinfo {author} {\bibfnamefont
  {L.}~\bibnamefont {Ricketson}}, \bibinfo {author} {\bibfnamefont
  {A.}~\bibnamefont {Scheinberg}}, \bibinfo {author} {\bibfnamefont {A.~Y.}\
  \bibnamefont {Sharma}},\ and\ \bibinfo {author} {\bibfnamefont {M.~E.}\
  \bibnamefont {Wall}},\ }\bibfield  {title} {\enquote {\bibinfo {title}
  {Enabling particle applications for exascale computing platforms},}\ }\href
  {https://doi.org/10.1177/10943420211022829} {\bibfield  {journal} {\bibinfo
  {journal} {Int. J. High Perform. Comput. Appl.}\ }\textbf {\bibinfo {volume}
  {35}},\ \bibinfo {pages} {572--597} (\bibinfo {year} {2021})}\BibitemShut
  {NoStop}%
\bibitem [{\citenamefont {Finkelstein}\ \emph {et~al.}(2021)\citenamefont
  {Finkelstein}, \citenamefont {Smith}, \citenamefont {Mniszewski},
  \citenamefont {Barros}, \citenamefont {Negre}, \citenamefont {Rubensson},\
  and\ \citenamefont {Niklasson}}]{JFinkelstein21}%
  \BibitemOpen
  \bibfield  {author} {\bibinfo {author} {\bibfnamefont {J.}~\bibnamefont
  {Finkelstein}}, \bibinfo {author} {\bibfnamefont {J.~S.}\ \bibnamefont
  {Smith}}, \bibinfo {author} {\bibfnamefont {S.~M.}\ \bibnamefont
  {Mniszewski}}, \bibinfo {author} {\bibfnamefont {K.}~\bibnamefont {Barros}},
  \bibinfo {author} {\bibfnamefont {C.~F.}\ \bibnamefont {Negre}}, \bibinfo
  {author} {\bibfnamefont {E.~H.}\ \bibnamefont {Rubensson}},\ and\ \bibinfo
  {author} {\bibfnamefont {A.~M.}\ \bibnamefont {Niklasson}},\ }\bibfield
  {title} {\enquote {\bibinfo {title} {Quantum-based molecular dynamics
  simulations using tensor cores},}\ }\href@noop {} {\bibfield  {journal}
  {\bibinfo  {journal} {J. Chem. Theory Comput.}\ }\textbf {\bibinfo {volume}
  {17}},\ \bibinfo {pages} {6180--6192} (\bibinfo {year} {2021})}\BibitemShut
  {NoStop}%
\bibitem [{\citenamefont {Pederson}\ \emph {et~al.}(2023)\citenamefont
  {Pederson}, \citenamefont {Kozlowski}, \citenamefont {Song}, \citenamefont
  {Beall}, \citenamefont {Ganahl}, \citenamefont {Hauru}, \citenamefont
  {Lewis}, \citenamefont {Yao}, \citenamefont {Mallick}, \citenamefont {Blum},\
  and\ \citenamefont {Vidal}}]{Pederson2023}%
  \BibitemOpen
  \bibfield  {author} {\bibinfo {author} {\bibfnamefont {R.}~\bibnamefont
  {Pederson}}, \bibinfo {author} {\bibfnamefont {J.}~\bibnamefont {Kozlowski}},
  \bibinfo {author} {\bibfnamefont {R.}~\bibnamefont {Song}}, \bibinfo {author}
  {\bibfnamefont {J.}~\bibnamefont {Beall}}, \bibinfo {author} {\bibfnamefont
  {M.}~\bibnamefont {Ganahl}}, \bibinfo {author} {\bibfnamefont
  {M.}~\bibnamefont {Hauru}}, \bibinfo {author} {\bibfnamefont {A.~G.~M.}\
  \bibnamefont {Lewis}}, \bibinfo {author} {\bibfnamefont {Y.}~\bibnamefont
  {Yao}}, \bibinfo {author} {\bibfnamefont {S.~B.}\ \bibnamefont {Mallick}},
  \bibinfo {author} {\bibfnamefont {V.}~\bibnamefont {Blum}},\ and\ \bibinfo
  {author} {\bibfnamefont {G.}~\bibnamefont {Vidal}},\ }\bibfield  {title}
  {\enquote {\bibinfo {title} {Large scale quantum chemistry with tensor
  processing units},}\ }\href {https://doi.org/10.1021/acs.jctc.2c00876}
  {\bibfield  {journal} {\bibinfo  {journal} {J. Chem. Theory Comput.}\
  }\textbf {\bibinfo {volume} {19}},\ \bibinfo {pages} {25--32} (\bibinfo
  {year} {2023})}\BibitemShut {NoStop}%
\bibitem [{\citenamefont {{Lupo Pasini}}\ \emph {et~al.}(2020)\citenamefont
  {{Lupo Pasini}}, \citenamefont {Turcksin}, \citenamefont {Ge},\ and\
  \citenamefont {Fattebert}}]{LUPOPASINI2020}%
  \BibitemOpen
  \bibfield  {author} {\bibinfo {author} {\bibfnamefont {M.}~\bibnamefont
  {{Lupo Pasini}}}, \bibinfo {author} {\bibfnamefont {B.}~\bibnamefont
  {Turcksin}}, \bibinfo {author} {\bibfnamefont {W.}~\bibnamefont {Ge}},\ and\
  \bibinfo {author} {\bibfnamefont {J.-L.}\ \bibnamefont {Fattebert}},\
  }\bibfield  {title} {\enquote {\bibinfo {title} {A parallel strategy for
  density functional theory computations on accelerated nodes},}\ }\href
  {https://doi.org/https://doi.org/10.1016/j.parco.2020.102703} {\bibfield
  {journal} {\bibinfo  {journal} {Parallel Computing}\ }\textbf {\bibinfo
  {volume} {100}},\ \bibinfo {pages} {102703} (\bibinfo {year}
  {2020})}\BibitemShut {NoStop}%
\bibitem [{cus()}]{cusolver}%
  \BibitemOpen
  \href@noop {} {\enquote {\bibinfo {title} {{cuSOLVER API Reference}},}\
  }\bibinfo {note} {\url{https://docs.nvidia.com/cuda/cusolver/}}\BibitemShut
  {NoStop}%
\bibitem [{\citenamefont {Paterson}\ and\ \citenamefont
  {Stockmeyer}(1973)}]{Patterson1973}%
  \BibitemOpen
  \bibfield  {author} {\bibinfo {author} {\bibfnamefont {M.~S.}\ \bibnamefont
  {Paterson}}\ and\ \bibinfo {author} {\bibfnamefont {L.~J.}\ \bibnamefont
  {Stockmeyer}},\ }\bibfield  {title} {\enquote {\bibinfo {title} {On the
  number of nonscalar multiplications necessary to evaluate polynomials},}\
  }\href {https://doi.org/10.1137/0202007} {\bibfield  {journal} {\bibinfo
  {journal} {SIAM Journal on Computing}\ }\textbf {\bibinfo {volume} {2}},\
  \bibinfo {pages} {60--66} (\bibinfo {year} {1973})}\BibitemShut {NoStop}%
\bibitem [{\citenamefont {Liang}\ \emph {et~al.}(2003)\citenamefont {Liang},
  \citenamefont {Saravanan}, \citenamefont {Shao}, \citenamefont {Baer},
  \citenamefont {Bell},\ and\ \citenamefont {Head-Gordon}}]{Liang2003}%
  \BibitemOpen
  \bibfield  {author} {\bibinfo {author} {\bibfnamefont {W.}~\bibnamefont
  {Liang}}, \bibinfo {author} {\bibfnamefont {C.}~\bibnamefont {Saravanan}},
  \bibinfo {author} {\bibfnamefont {Y.}~\bibnamefont {Shao}}, \bibinfo {author}
  {\bibfnamefont {R.}~\bibnamefont {Baer}}, \bibinfo {author} {\bibfnamefont
  {A.~T.}\ \bibnamefont {Bell}},\ and\ \bibinfo {author} {\bibfnamefont
  {M.}~\bibnamefont {Head-Gordon}},\ }\bibfield  {title} {\enquote {\bibinfo
  {title} {Improved {Fermi} operator expansion methods for fast electronic
  structure calculations},}\ }\href@noop {} {\bibfield  {journal} {\bibinfo
  {journal} {J. Chem. Phys.}\ }\textbf {\bibinfo {volume} {119}},\ \bibinfo
  {pages} {4117--4125} (\bibinfo {year} {2003})}\BibitemShut {NoStop}%
\bibitem [{\citenamefont {Liang}\ \emph {et~al.}(2004)\citenamefont {Liang},
  \citenamefont {Baer}, \citenamefont {Saravanan}, \citenamefont {Shao},
  \citenamefont {Bell},\ and\ \citenamefont {Head-Gordon}}]{Liang2004}%
  \BibitemOpen
  \bibfield  {author} {\bibinfo {author} {\bibfnamefont {W.~Z.}\ \bibnamefont
  {Liang}}, \bibinfo {author} {\bibfnamefont {R.}~\bibnamefont {Baer}},
  \bibinfo {author} {\bibfnamefont {C.}~\bibnamefont {Saravanan}}, \bibinfo
  {author} {\bibfnamefont {Y.}~\bibnamefont {Shao}}, \bibinfo {author}
  {\bibfnamefont {A.~T.}\ \bibnamefont {Bell}},\ and\ \bibinfo {author}
  {\bibfnamefont {M.}~\bibnamefont {Head-Gordon}},\ }\bibfield  {title}
  {\enquote {\bibinfo {title} {Fast methods for resumming matrix polynomials
  and {Chebyshev} matrix polynomials},}\ }\href@noop {} {\bibfield  {journal}
  {\bibinfo  {journal} {J. Comp. Phys.}\ }\textbf {\bibinfo {volume} {194}},\
  \bibinfo {pages} {575--587} (\bibinfo {year} {2004})}\BibitemShut {NoStop}%
\bibitem [{mag()}]{magma}%
  \BibitemOpen
  \href@noop {} {\enquote {\bibinfo {title} {{MAGMA: Matrix Algebra on GPU and
  Multicore Architectures}},}\ }\bibinfo {note}
  {\url{https://icl.utk.edu/magma/index.html}}\BibitemShut {NoStop}%
\bibitem [{\citenamefont {Demmel}(1997)}]{demmel97}%
  \BibitemOpen
  \bibfield  {author} {\bibinfo {author} {\bibfnamefont {J.~W.}\ \bibnamefont
  {Demmel}},\ }\href {https://doi.org/10.1137/1.9781611971446} {\emph {\bibinfo
  {title} {Applied Numerical Linear Algebra}}}\ (\bibinfo  {publisher} {SIAM},\
  \bibinfo {year} {1997})\BibitemShut {NoStop}%
\bibitem [{ope()}]{openblas}%
  \BibitemOpen
  \href@noop {} {\enquote {\bibinfo {title} {{OpenBLAS}, an optimized {BLAS}
  library},}\ }\bibinfo {note} {\url{http://www.openblas.net}}\BibitemShut
  {NoStop}%
\bibitem [{\citenamefont {Dongarra}\ \emph {et~al.}(2014)\citenamefont
  {Dongarra}, \citenamefont {Gates}, \citenamefont {Haidar}, \citenamefont
  {Kurzak}, \citenamefont {Luszczek}, \citenamefont {Tomov},\ and\
  \citenamefont {Yamazaki}}]{magma2014}%
  \BibitemOpen
  \bibfield  {author} {\bibinfo {author} {\bibfnamefont {J.}~\bibnamefont
  {Dongarra}}, \bibinfo {author} {\bibfnamefont {M.}~\bibnamefont {Gates}},
  \bibinfo {author} {\bibfnamefont {A.}~\bibnamefont {Haidar}}, \bibinfo
  {author} {\bibfnamefont {J.}~\bibnamefont {Kurzak}}, \bibinfo {author}
  {\bibfnamefont {P.}~\bibnamefont {Luszczek}}, \bibinfo {author}
  {\bibfnamefont {S.}~\bibnamefont {Tomov}},\ and\ \bibinfo {author}
  {\bibfnamefont {I.}~\bibnamefont {Yamazaki}},\ }\bibfield  {title} {\enquote
  {\bibinfo {title} {Accelerating numerical dense linear algebra calculations
  with {GPUs}},}\ }\href@noop {} {\bibfield  {journal} {\bibinfo  {journal}
  {Numerical Computations with GPUs}\ ,\ \bibinfo {pages} {1--26}} (\bibinfo
  {year} {2014})}\BibitemShut {NoStop}%
\bibitem [{Note1()}]{Note1}%
  \BibitemOpen
  \bibinfo {note} {It is worth noting that ${\protect \tt magma\protect
  \_dgemm}$ and ${\protect \tt magmablas\protect \_dgemm}$ are different
  implementations within MAGMA, with the former utilizing vendor libraries and
  latter the intrinsic MAGMA version. In our implementations, we use ${\protect
  \tt magma\protect \_dgemm}$}\BibitemShut {NoStop}%
\bibitem [{\citenamefont {Brown}\ \emph {et~al.}(2020)\citenamefont {Brown},
  \citenamefont {Abdelfattah}, \citenamefont {Tomov},\ and\ \citenamefont
  {Dongarra}}]{magma2020}%
  \BibitemOpen
  \bibfield  {author} {\bibinfo {author} {\bibfnamefont {C.}~\bibnamefont
  {Brown}}, \bibinfo {author} {\bibfnamefont {A.}~\bibnamefont {Abdelfattah}},
  \bibinfo {author} {\bibfnamefont {S.}~\bibnamefont {Tomov}},\ and\ \bibinfo
  {author} {\bibfnamefont {J.}~\bibnamefont {Dongarra}},\ }\bibfield  {title}
  {\enquote {\bibinfo {title} {Design, optimization, and benchmarking of dense
  linear algebra algorithms on {AMD} {GPU}s},}\ }in\ \href
  {https://doi.org/10.1109/HPEC43674.2020.9286214} {\emph {\bibinfo {booktitle}
  {2020 IEEE High Performance Extreme Computing Conference (HPEC)}}}\ (\bibinfo
  {year} {2020})\ pp.\ \bibinfo {pages} {1--7}\BibitemShut {NoStop}%
\bibitem [{\citenamefont {Bock}\ \emph {et~al.}(2018)\citenamefont {Bock},
  \citenamefont {Negre}, \citenamefont {Mniszewski}, \citenamefont
  {Mohd-Yusof}, \citenamefont {Aradi}, \citenamefont {Fattebert}, \citenamefont
  {Osei-Kuffuor}, \citenamefont {Germann},\ and\ \citenamefont
  {Niklasson}}]{Bock2018}%
  \BibitemOpen
  \bibfield  {author} {\bibinfo {author} {\bibfnamefont {N.}~\bibnamefont
  {Bock}}, \bibinfo {author} {\bibfnamefont {C.~F.~A.}\ \bibnamefont {Negre}},
  \bibinfo {author} {\bibfnamefont {S.~M.}\ \bibnamefont {Mniszewski}},
  \bibinfo {author} {\bibfnamefont {J.}~\bibnamefont {Mohd-Yusof}}, \bibinfo
  {author} {\bibfnamefont {B.}~\bibnamefont {Aradi}}, \bibinfo {author}
  {\bibfnamefont {J.-L.}\ \bibnamefont {Fattebert}}, \bibinfo {author}
  {\bibfnamefont {D.}~\bibnamefont {Osei-Kuffuor}}, \bibinfo {author}
  {\bibfnamefont {T.~C.}\ \bibnamefont {Germann}},\ and\ \bibinfo {author}
  {\bibfnamefont {A.~M.~N.}\ \bibnamefont {Niklasson}},\ }\bibfield  {title}
  {\enquote {\bibinfo {title} {The basic matrix library ({BML}) for quantum
  chemistry},}\ }\href@noop {} {\bibfield  {journal} {\bibinfo  {journal} {J.
  Supercomput.}\ }\textbf {\bibinfo {volume} {74}},\ \bibinfo {pages}
  {6201--6219} (\bibinfo {year} {2018})}\BibitemShut {NoStop}%
\bibitem [{\citenamefont {Niklasson}\ \emph {et~al.}(2022)\citenamefont
  {Niklasson}, \citenamefont {Mniszewski}, \citenamefont {Negre}, \citenamefont
  {Wall}, \citenamefont {Cawkwell},\ and\ \citenamefont {Bock}}]{2022progress}%
  \BibitemOpen
  \bibfield  {author} {\bibinfo {author} {\bibfnamefont {A.~M.}\ \bibnamefont
  {Niklasson}}, \bibinfo {author} {\bibfnamefont {S.~M.}\ \bibnamefont
  {Mniszewski}}, \bibinfo {author} {\bibfnamefont {C.~F.~A.}\ \bibnamefont
  {Negre}}, \bibinfo {author} {\bibfnamefont {M.~E.}\ \bibnamefont {Wall}},
  \bibinfo {author} {\bibfnamefont {M.~J.}\ \bibnamefont {Cawkwell}},\ and\
  \bibinfo {author} {\bibfnamefont {N.}~\bibnamefont {Bock}},\ }\href@noop {}
  {\enquote {\bibinfo {title} {{PROGRESS}, version 1.2},}\ } (\bibinfo {year}
  {2022}),\ \bibinfo {note}
  {\url{https://github.com/lanl/qmd-progress}}\BibitemShut {NoStop}%
\bibitem [{\citenamefont {Hourahine}\ \emph {et~al.}(2020)\citenamefont
  {Hourahine}, \citenamefont {Aradi}, \citenamefont {Blum}, \citenamefont
  {Bonaf{\'e}}, \citenamefont {Buccheri}, \citenamefont {Camacho},
  \citenamefont {Cevallos}, \citenamefont {Deshaye}, \citenamefont
  {Dumitric{\u{a}}}, \citenamefont {Dominguez} \emph {et~al.}}]{dftb+}%
  \BibitemOpen
  \bibfield  {author} {\bibinfo {author} {\bibfnamefont {B.}~\bibnamefont
  {Hourahine}}, \bibinfo {author} {\bibfnamefont {B.}~\bibnamefont {Aradi}},
  \bibinfo {author} {\bibfnamefont {V.}~\bibnamefont {Blum}}, \bibinfo {author}
  {\bibfnamefont {F.}~\bibnamefont {Bonaf{\'e}}}, \bibinfo {author}
  {\bibfnamefont {A.}~\bibnamefont {Buccheri}}, \bibinfo {author}
  {\bibfnamefont {C.}~\bibnamefont {Camacho}}, \bibinfo {author} {\bibfnamefont
  {C.}~\bibnamefont {Cevallos}}, \bibinfo {author} {\bibfnamefont
  {M.}~\bibnamefont {Deshaye}}, \bibinfo {author} {\bibfnamefont
  {T.}~\bibnamefont {Dumitric{\u{a}}}}, \bibinfo {author} {\bibfnamefont
  {A.}~\bibnamefont {Dominguez}}, \emph {et~al.},\ }\bibfield  {title}
  {\enquote {\bibinfo {title} {{DFTB+}, a software package for efficient
  approximate density functional theory based atomistic simulations},}\
  }\href@noop {} {\bibfield  {journal} {\bibinfo  {journal} {J. Chem. Phys.}\
  }\textbf {\bibinfo {volume} {152}},\ \bibinfo {pages} {124101} (\bibinfo
  {year} {2020})}\BibitemShut {NoStop}%
\bibitem [{\citenamefont {Bock}\ \emph {et~al.}(2023)\citenamefont {Bock},
  \citenamefont {Cawkwell}, \citenamefont {Coe}, \citenamefont {Krishnapriyan},
  \citenamefont {Kroonblawd}, \citenamefont {Lang}, , \citenamefont {Liu},
  \citenamefont {Saez}, \citenamefont {Mniszewski}, \citenamefont {Negre},
  \citenamefont {Niklasson}, \citenamefont {Sanville}, \citenamefont {Wood},\
  and\ \citenamefont {Yang}}]{latte}%
  \BibitemOpen
  \bibfield  {author} {\bibinfo {author} {\bibfnamefont {N.}~\bibnamefont
  {Bock}}, \bibinfo {author} {\bibfnamefont {M.~J.}\ \bibnamefont {Cawkwell}},
  \bibinfo {author} {\bibfnamefont {J.~D.}\ \bibnamefont {Coe}}, \bibinfo
  {author} {\bibfnamefont {A.}~\bibnamefont {Krishnapriyan}}, \bibinfo {author}
  {\bibfnamefont {M.~P.}\ \bibnamefont {Kroonblawd}}, \bibinfo {author}
  {\bibfnamefont {A.}~\bibnamefont {Lang}}, , \bibinfo {author} {\bibfnamefont
  {C.}~\bibnamefont {Liu}}, \bibinfo {author} {\bibfnamefont {E.~M.}\
  \bibnamefont {Saez}}, \bibinfo {author} {\bibfnamefont {S.~M.}\ \bibnamefont
  {Mniszewski}}, \bibinfo {author} {\bibfnamefont {C.~F.~A.}\ \bibnamefont
  {Negre}}, \bibinfo {author} {\bibfnamefont {A.~M.~N.}\ \bibnamefont
  {Niklasson}}, \bibinfo {author} {\bibfnamefont {E.}~\bibnamefont {Sanville}},
  \bibinfo {author} {\bibfnamefont {M.~A.}\ \bibnamefont {Wood}},\ and\
  \bibinfo {author} {\bibfnamefont {P.}~\bibnamefont {Yang}},\ }\href@noop {}
  {\enquote {\bibinfo {title} {{LATTE}: Developer repository for the {LATTE}
  code},}\ } (\bibinfo {year} {2023}),\ \bibinfo {note}
  {\url{https://github.com/lanl/LATTE}}\BibitemShut {NoStop}%
\bibitem [{\citenamefont {Soler}\ \emph {et~al.}(2002)\citenamefont {Soler},
  \citenamefont {Artacho}, \citenamefont {Gale}, \citenamefont {Garc{\'\i}a},
  \citenamefont {Junquera}, \citenamefont {Ordej{\'o}n},\ and\ \citenamefont
  {S{\'a}nchez-Portal}}]{Soler2002-wc}%
  \BibitemOpen
  \bibfield  {author} {\bibinfo {author} {\bibfnamefont {J.~M.}\ \bibnamefont
  {Soler}}, \bibinfo {author} {\bibfnamefont {E.}~\bibnamefont {Artacho}},
  \bibinfo {author} {\bibfnamefont {J.~D.}\ \bibnamefont {Gale}}, \bibinfo
  {author} {\bibfnamefont {A.}~\bibnamefont {Garc{\'\i}a}}, \bibinfo {author}
  {\bibfnamefont {J.}~\bibnamefont {Junquera}}, \bibinfo {author}
  {\bibfnamefont {P.}~\bibnamefont {Ordej{\'o}n}},\ and\ \bibinfo {author}
  {\bibfnamefont {D.}~\bibnamefont {S{\'a}nchez-Portal}},\ }\bibfield  {title}
  {\enquote {\bibinfo {title} {The {SIESTA} method for ab initio order-{N}
  materials simulation},}\ }\href@noop {} {\bibfield  {journal} {\bibinfo
  {journal} {J. Phys. Condens. Matter}\ }\textbf {\bibinfo {volume} {14}},\
  \bibinfo {pages} {2745} (\bibinfo {year} {2002})}\BibitemShut {NoStop}%
\end{thebibliography}%

\end{document}